\documentclass[12pt,3p,times]{elsarticle}
\usepackage{color,graphicx}
\usepackage{color,graphicx,amsmath,amssymb}

\def\slash#1{\mbox{$\not \!\! #1$}}
\def\Dslash{ {\slash\,{\cal D}}}

\def\lvec#1{\setbox0=\hbox{$#1$}
    \setbox1=\hbox{$\scriptstyle\leftarrow$}
    #1\kern-\wd0\smash{
    \raise\ht0\hbox{$\raise1pt\hbox{$\scriptstyle\leftarrow$}$}}
    \kern-\wd1\kern\wd0}
\def\rvec#1{\setbox0=\hbox{$#1$}
    \setbox1=\hbox{$\scriptstyle\rightarrow$}
    #1\kern-\wd0\smash{
    \raise\ht0\hbox{$\raise1pt\hbox{$\scriptstyle\rightarrow$}$}}
    \kern-\wd1\kern\wd0}

\def\diracstar#1#2{
    \setbox0=\hbox{$\gamma$}\setbox1=\hbox{$\gamma_{#1}$}\includegraphics[]{QNEW_36-eps-converted-to.pdf}

    \gamma_{#1}\kern-\wd1\kern\wd0
    \smash{\raise4.5pt\hbox{$\scriptstyle#2$}}}

\def\tr{\,\hbox{tr}\,}

\newcommand{\beq}{\begin{equation}}
\newcommand{\eeq}{\end{equation}}
\newcommand{\beqn}{\begin{eqnarray}}
\newcommand{\eeqn}{\end{eqnarray}}
\newcommand{\nn}{\nonumber}

\renewcommand\appendix{\par
  \setcounter{section}{0}%
  \setcounter{subsection}{0}%
  \setcounter{equation}{0}
 \gdef\thefigure{\@Alph\c@section.\arabic{figure}}%
 \gdef\thetable{\@Alph\c@section.\arabic{table}}%
 \gdef\thesection{\appendixname~\@Alph\c@section}\@addtoreset{equation}{section}%
\gdef\theequation{\@Alph\c@section.\arabic{equation}}%
  \addtocontents{toc}{\string\let\string\numberline\string\tmptocnumberline}{}{}
}

\usepackage{chir_layout}

\begin{document}
\begin{frontmatter}
\title{{
\bf A road to an elementary particle physics model \\ with no Higgs~-~I }
\vspace*{-4mm}}

\author{G.C.\ Rossi}

\address{  {\small Dipartimento di Fisica, Universit\`a di  Roma
  ``{\it Tor Vergata}'' \\ INFN, Sezione di Roma  ``{\it Tor Vergata}'' }\\
  {\small Via della Ricerca Scientifica - 00133 Roma, Italy}\\
  {\small Centro Fermi - Museo Storico della Fisica e Centro Studi e Ricerche E.\ Fermi, \\ Via Panisperna 89a, 00184 Roma, Italy}}


\vspace*{-3mm}
\begin{abstract}

This is the first of two companion papers in which we prove that the recently discovered non-perturbative mechanism capable of giving mass to elementary fermions can also generate a mass for the electro-weak bosons, when weak interactions are switched on, and can thus be used as a viable alternative to the Higgs scenario.\ It can be demonstrated that the non-perturbatively generated fermion and $W$ masses have the expression $m_{f}\sim C_f(\alpha)\Lambda_{RGI}$ and $M_W\sim g_w c_w(\alpha)\Lambda_{RGI}$, respectively, where $C_f(\alpha)$ and $c_w(\alpha)$ are functions of the gauge couplings, $g_w$ is the weak coupling and $\Lambda_{RGI}$ the RGI scale of the theory. In view of this parametric structure we are led to argue that to get top quark and $W$ boson masses of the correct order of magnitude, a realistic model must include a yet unobserved sector of massive fermions (Tera-fermions) subjected, besides Standard Model interactions, to some kind of super-strong gauge interactions (Tera-interactions) so that the full theory (including Standard Model and Tera-particles) will have an RGI scale (we call it $\Lambda_T$) much larger than $\Lambda_{QCD}$, located in the few TeV region. The further extension of the model with the introduction of hypercharge interactions and particles singlets under strong interactions (leptons and Tera-leptons) is the object of the companion paper, where we also discuss some interesting phenomenological implications of the scenario we are advocating. As for the relation of our approach to the Standard Model, we show that, upon integrating out the (heavy) Tera-degrees of freedom, the resulting low energy effective Lagrangian closely resembles the Standard Model Lagrangian. The argument rests on the key conjecture that the 125~GeV resonance detected at LHC is a $W^+W^-/ZZ$ composite state, bound by Tera-particle exchanges, and not an elementary particle. 

Though limited in its scope (for the moment we restrict to the case of only one family, neglecting weak isospin splitting), the present approach has a certain number of merits with respect to the Standard Model. Lacking a fundamental Higgs, it offers a radical solution of the Higgs mass tuning problem. It provides a physical interpretation of the electro-weak scale, with the latter being (a fraction of) the scale, $\Lambda_T$, of a new interaction. It helps reducing the number of the Standard Model parameters, as elementary particle masses are non-perturbatively fully determined by the dynamics of the theory.
\end{abstract}
\end{frontmatter}

\section{Introduction}
\label{sec:INTRO}


The authors of ref.~\cite{Frezzotti:2014wja} introduced an interesting field theoretical renormalizable model where an SU(2) fermion doublet, subjected to non-abelian gauge interactions of the QCD type, is coupled to a complex scalar field via a $d = 4$ Yukawa term and an ``irrelevant'' $d > 4$ Wilson-like operator~\cite{WilsonLQCD}. Notwithstanding the fact that both terms break chiral invariance, it was shown in~\cite{Frezzotti:2014wja}, and numerically checked in~\cite{Capitani:2019syo} in a set of dedicated lattice simulations, that there exists a critical value of the Yukawa coupling where chiral symmetry is recovered, up to corrections formally vanishing as the UV cut-off is removed. 

Although an exact $\chi_L\times\chi_R$ symmetry, that rotates all the matter fields of the Lagrangian, forbids the existence of standard fermionic mass operators, i.e.\ operators of the form $m\bar \psi \psi$, it can be shown that in the Nambu--Goldstone (NG) phase of the critical theory where the $\chi_L\times\chi_R$ symmetry is spontaneously broken, peculiar non-perturbative (NP) mass terms emerge of the kind $C\Lambda_{\rm RGI}(\bar \psi_LU\psi_R+\bar \psi_RU^\dagger\psi_L)$, where $U=\exp{i\vec\tau \vec \zeta/C\Lambda_{\rm RGI}}$ is the exponential of the Goldstone bosons associated with the spontaneously broken $\chi_L\times\chi_R$ symmetry, $\Lambda_{\rm RGI}$ is the RGI scale of the theory and $C$ is a computable coefficient, function of the gauge couplings. Upon expanding $U$ around the identity we see that NP O($\Lambda_{\rm RGI}$) masses for the elementary fermions get dynamically generated (see below eq.~(\ref{EQ1})) plus a wealth of non-linear NG bosons interactions.

\subsection{NP mass generation mechanism}

NP masses emerge as a consequence of a sort of ``interference'' between  residual UV chiral breaking effects left behind at the critical value of the Yukawa coupling in the NG phase of the theory, and IR features triggered by the phenomenon of spontaneous breaking of (the recovered) chiral symmetry which standardly takes place in a strongly interacting theory. 

A detailed analysis of this subtle field theoretical interplay shows that the non perturbatively (NP-ly) generated elementary fermion masses have the parametric expression~\cite{Frezzotti:2014wja,Frezzotti:2013raa}  
\begin{equation}
m_{f}\sim C_f(\alpha) \Lambda_{\rm RGI} \, ,\label{EQ1}
\end{equation}
where the coefficient $C_f(\alpha)$ is a function of the gauge coupling. If we take the ``irrelevant'' chiral breaking Wilson-like term to be a $d=6$ operator (as we do for illustrative purposes in eqs.~(\ref{SULL}) and~(\ref{SULLWQ}) below), one finds at the lowest loop order $C_f(\alpha) ={\mbox{O}}(\alpha^2)$. 

A far reaching phenomenological implication of the formula~(\ref{EQ1}), when referred to the top quark, is that the relation $m_{\rm top}\sim{\mbox{O}}(\Lambda_{\rm RGI})$ requires $\Lambda_{\rm RGI}\gg\Lambda_{QCD}$ and of the order of a few TeVs if eq.~(\ref{EQ1}) has to reproduce the correct (order of magnitude of the) top mass experimental value. We are thus led to conjecture that a new sector of super-strongly interacting particles~\footnote{To avoid any misunderstanding we explicitly remark that here "super" has nothing to do with super-symmetry.}, gauge-invariantly coupled to standard matter, needs to exist so that the complete theory, encompassing the new and the Standard Model (SM) particles, will have an RGI scale in the TeV region~\cite{Frezzotti:2018zsy,Frezzotti:2019npa,Rossi:2022vlr,Rossi:2022xpa,Rossi:2023jjv}. 

We want to immediately remark that the need for assuming the existence of a super-strongly interacting sector here is totally different from the reason why Technicolor was introduced in refs.~\cite{Susskind:1978ms,Weinberg:1979bn}. Technicolor was invoked as a way to provide mass to the electroweak (EW) bosons in the first place, while in the present approach super-strong interactions need to be introduced to get the right order of magnitude of the dynamically generated top quark mass and, as we will show in sect.~\ref{sec:EEPM}, also of the $W$ mass. 

As we have said above, a key feature of our model is that (irrespective of the value of the Yukawa coupling) the Lagrangian enjoys an exact symmetry, $\chi_L\times\chi_R$, protecting elementary particle masses against power divergent quantum corrections, unlike what happens in Wilson lattice QCD (WLQCD) where the fermion mass is affected by a linearly divergent power correction~\cite{WilsonLQCD}.\ At variance with Technicolor, the same exact symmetry makes all fermion--anti-fermion vev's (condensates) vanishing, thus precluding the possibility of using them to generate masses.

In any case, as suggested by Glashow~\cite{Glashow:2005jy}, in the following to avoid confusion, we will refer to these new super-strong interactions as Tera-interactions and to the new set of particles as Tera-particles.

In the present paper we show that the model proposed in~\cite{Frezzotti:2014wja} can be naturally extended to incorporate weak interactions and Tera-particles (see refs.~\cite{Frezzotti:2018zsy,Frezzotti:2019npa} for an early formulation of the extended model). The $W$ bosons as well as the Tera-fermions will acquire a NP mass O($\Lambda_{\rm RGI}$) via the same mechanism that leads to eq.~(\ref{EQ1}). For the $W$ mass one gets (see sect.~\ref{sec:WMASS}) 
\begin{equation}
M_W\sim g_w c_w(\alpha)\Lambda_{\rm RGI}\, ,
\label{EQ2}
\end{equation} 
where $g_w$ is the weak coupling and $\alpha$ is a short-hand for the set of gauge couplings $[\alpha_w,\alpha_s, \alpha_T]$ with $\alpha_w$, $\alpha_s$ and $\alpha_T$ referring to weak, strong and Tera-strong gauge interactions, respectively. In the following we will denote by $\Lambda_{T}$ the RGI scale of the whole theory where the subscript $T$ is to remind us that we are including Tera-particles in the theory. The lowest loop order parametric expression of $c_w(\alpha)$ and that of the NP Tera-fermion mass are derived in sect.~\ref{sec:NPM}. 

In this first paper we limit ourselves to discuss how the model of ref.~\cite{Frezzotti:2014wja} can be extended to incorporate weak interactions and Tera-particles. In the companion paper~\cite{Rossi:2023wok} (hereafter referred to as~(II)), besides taking up a number of technical issues left aside here, we address the fundamental issue of adding hypercharge interactions and particles, singlets under strong interactions (leptons and Tera-leptons), and we present some interesting phenomenological implications of the model. 

\subsection{Comparing with the Standard Model}
\label{sec:CWSM}

Eqs.~(\ref{EQ1})-(\ref{EQ2}) are very similar to the expression of the SM Higgs-like mass of fermions and $W$'s, respectively, with however two fundamental differences. The first is that the scale of the masses is not the vev of the Higgs field (whose value in the SM is, for instance, determined by fitting the $W$ mass), but a dynamical quantity associated to a new interaction. The second, related difference is that the values of the Yukawa couplings, which in the SM are tuned by hand to match the phenomenological values of quark masses (and, if present, also of lepton masses, see~(II)) in the present approach are not free parameters. They are in principle calculable and are determined by the dynamics of the theory. 

From the above discussion we conclude that the general NP scenario for elementary particle mass generation we are describing in this investigation can be considered as a valid alternative to the Higgs mechanism, with the extra advantage that lacking a fundamental Higgs, we will not have to worry about the fine tuning problem of its mass. Though we may need to reassess the question of the stability of the EW vacuum~\cite{Cabibbo:1979ay,Ellis:2009tp,Alekhin:2012py,Salvio:2015cja}. A further conceptual bonus is that we have a natural interpretation of the magnitude of the electroweak scale (EW) scale, as (a fraction of the) the physical dynamical parameter, $\Lambda_T$.

\subsubsection{A few remarks}

There is a number of additional interesting features and implications of the approach we are advocating in this work that are worth mentioning. 

First of all, lacking the need for the existence of a fundamental Higgs boson for elementary particle mass generation, we are in the obligation of finding a convincing interpretation for the 125~GeV resonance identified at LHC~\cite{Aad:2012tfa,Chatrchyan:2012xdj}. We conjecture that this particle is not a fundamental object. We propose to interpret it as a $W^+W^-/ZZ$ composite state, bound by exchanges of Tera-particles which, being charged under EW interactions, can couple to the EW bosons. If the above interpretation is correct, this scalar particle must be incorporated in the low energy effective Lagrangian (LEEL) of the theory that one gets by integrating out the heavy Tera-degrees of freedom (Tera-dof's) since its experimental mass is much smaller than the $\Lambda_T$ scale. Not unexpectedly, at (momenta)$^2 \ll \Lambda_T^2$ the resulting $d=4$ part of the LEEL of the model is seen to resemble very much the SM Lagrangian. We shall discuss this issue in detail in sect.~\ref{sec:ITDF}. 

Secondly, as it was shown in ref.~\cite{Frezzotti:2016bes}, with a reasonable choice of the elementary particle content, a realistic theory extending the SM with the inclusion of the new Tera-sector leads to gauge coupling unification at a $\sim 10^{18}$~GeV scale. This somewhat large unification scale is likely to yield a proton life-time comfortably larger than the present bound $\tau_{\rm prot} > 1.7 \times10^{34}$ years~\cite{Bajc:2016qcc}. 

We end by observing that, as discussed in sect.~\ref{sec:UNIV}, the problem of understanding the multiplicity of families (and possibly cope with weak isospin splitting) is related to the questions of the universality of the NP mass generation mechanism we are describing in this work, and the predicting  power of the present approach. Both issues are under active investigation. 

\subsection{Outline of the paper}

The outline of this first paper is as follows. In sect.~\ref{sec:STM} for completeness and to set our notations we review the key features of the model introduced in ref.~\cite{Frezzotti:2014wja}. In sect.~\ref{sec:WMASS} we discuss how to extend it with the introduction of weak and Tera-interactions. In sect.~\ref{sec:ITDF} we prove that the LEEL one obtains by integrating out the heavy Tera-dof's closely resembles to the SM Lagrangian. In sect.~\ref{sec:UNIV} we touch the issue of universality and predictive power. A few conclusions and an outlook of our future lines of investigation can be found in sect.~\ref{sec:CAO}. In Appendix~A we recall the procedure that leads to the equations determining the values of the Lagrangian parameters, which specify the critical theory endowed with an enlarged symmetry of chiral nature. In Appendix~B we justify the form that the quantum effective Lagrangian (QEL) functional of the theory~\footnote{By QEL we mean the generating functional of the 1PI vertices from which one can directly extract the full quantum information of the model. To be precise in this paper we make a distinction between the QEL functional that represents the Legendre transform of the partition function and the LEEL of the theory, valid at small momenta, that is obtained upon integrating out all the dof's heavier than the chosen momentum scale.} takes in the NG phase at the critical point. In Appendix~\ref{sec:APPC} we discuss the form of the effective scalar propagator in the critical limit. The way in which the transversality property of the $W$ polarization amplitude is implemented in the present NP approach to mass generation is explained in Appendix~\ref{sec:APPD}. 

\section{A toy-model} 
\label{sec:STM} 

For the reader's convenience we summarize in this section the results of ref.~\cite{Frezzotti:2014wja}. The simplest situation in which the NP mass generation mechanism outlined in the Introduction takes place is realized in a field theory where an SU(2) fermion doublet, subjected to non-abelian gauge interactions of the QCD type, is coupled to a complex scalar field via a $d=4$ Yukawa term and an ``irrelevant'' $d=6$ Wilson-like operator~\cite{WilsonLQCD}, which both break chiral invariance. The Lagrangian of this ``toy-model'' reads 
\begin{eqnarray}
\hspace{-1.4cm}&&{\cal L}_{\rm toy}(q,A;\Phi)= {\cal L}_{K}(q,A;\Phi)+{\cal V}(\Phi)
+ {\cal L}_{Yuk}(q;\Phi)+{\cal L}_{Wil}(q,A;\Phi) \label{SULL}\\
\hspace{-1.4cm}&&\quad\bullet\,{\cal L}_{K}(q,A;\Phi)= \frac{1}{4}F^A\!\cdot\! F^A+\big{(} \bar q_L \,\Dslash^A q_L+\bar q_R\, \Dslash^A q_R\big{)} +\frac{1}{2}{\tr}\big{[}\partial_\mu\Phi^\dagger\partial_\mu\Phi\big{]}\label{LKIN}\\
\hspace{-1.4cm}&&\quad\bullet\,{\cal V}(\Phi)= \frac{\mu_0^2}{2}{\tr}\big{[}\Phi^\dagger\Phi\big{]}+\frac{\lambda_0}{4}\big{(}{\tr}\big{[}\Phi^\dagger\Phi\big{]}\big{)}^2\label{VPHI}\\
\hspace{-1.4cm}&&\quad\bullet\,{\cal L}_{Yuk}(q;\Phi)= \eta\,\big{(} \bar q_L\Phi q_R+\bar q_R \Phi^\dagger q_L\big{)} \label{LYUK}\\
\hspace{-1.4cm}&&\quad\bullet\,{\cal L}_{Wil}(q,A;\Phi)= \frac{b^2}{2}\rho\,\big{(}\bar q_L\overleftarrow{\cal D}^A_\mu\Phi {\cal D}^A_\mu q_R+\bar q_R \overleftarrow{\cal D}^A_\mu \Phi^\dagger {\cal D}^A_\mu q_L\big{)}\label{LWIL}\, ,
\end{eqnarray}
where $q_L=(u_L,d_L)^T$ and $q_R=(u_R,d_R)^T$ are fermion iso-doublets and $\Phi = \varphi_01\!\!1+i \varphi_j\tau^j= [-i\tau_2 \varphi^\star| \,\varphi]$ is a $2\times 2$ matrix with $\varphi=(\varphi_2-i\varphi_1,\varphi_0-i\varphi_3)^T$ a complex scalar doublet, singlet under color SU($N_c$) gauge transformations. The length scale $b$ is the inverse of the UV cutoff, $b^{-1} \sim \Lambda_{UV}$, $\eta$ is the Yukawa coupling, $\rho$ is introduced to keep track of ${\cal L}_{Wil}$ and ${\cal D}_\mu^A$ is the gauge-covariant derivative. 

A few remarks are in order here.\ First of all, we want to stress that the  distinctive feature of the Lagrangian~(\ref{SULL}) is the presence of the ``irrelevant'' $d>4$ chiral breaking Wilson-like term. Secondly, because of its presence we have immediately introduced a Yukawa term in the fundamental Lagrangian as it will be anyway generated by loop corrections. Finally, we observe that, despite the presence of the $d=6$ ${\cal L}_{Wil}$ term, ${\cal L}_{\rm toy}$ is power counting renormalizable, because the $d=6$ Wilson-like operator appears in the Lagrangian multiplied by two inverse powers of the UV cutoff. It is worth recalling that a similar situation occurs in WLQCD where the $d=5$ Wilson term in the lattice action is multiplied by one power of the lattice spacing, thus making the lattice action renormalizable and amenable to Monte~Carlo simulations~\cite{WilsonLQCD,Sint:2000vc}. 

\subsection{Symmetries}
\label{sec:SYMME}

Among other obvious symmetries, ${\cal L}_{\rm toy}$ is invariant under the (global) transformations $\chi_L\times \chi_R$ involving fermions and scalars given by ($\Omega_{L/R} \in {\mbox{SU}}(2)$) 
\begin{eqnarray}
\hspace{-.5cm}&&\chi_L\times \chi_R =  [\tilde\chi_L\times (\Phi\to\Omega_L\Phi)]\times [\tilde\chi_R\times (\Phi\to\Phi\Omega_R^\dagger)] \label{CHIL}\\
\hspace{-.5cm}&&\qquad\tilde\chi_{L} : q_{L}\rightarrow\Omega_{L} q_{L} \, ,\qquad\qquad \,\bar q_{L}\rightarrow \bar q_{L}\Omega_{L}^\dagger \label{GTWTL}\\
\hspace{-.5cm}&&\qquad\tilde\chi_{R} : q_{R}\rightarrow\Omega_{R} q_{R} \, ,\qquad \qquad\bar q_{R}\rightarrow \bar q_{R}\Omega_{R}^\dagger \label{GTWTR}
\end{eqnarray} 
The exact $\chi_L\times \chi_R$ symmetry can be realized either {\it \'a la} Wigner or {\it \'a la} NG depending on the shape of the scalar potential, ${\cal V}(\Phi)$. In any case, as we mentioned in the Introduction, no linearly divergent fermion mass can be generated by quantum corrections because the mass operator $\bar q_L q_R+ \bar q_R q_L$ is not invariant under $\chi_L\times\chi_R$. Obviously, this also means that no fermion--anti-fermion vev's of the kind $\langle \bar q_L q_R+ \bar q_R q_L\rangle$ (condensates) can ever be non-vanishing. 

\subsection{Chiral invariance}
\label{sec:CHIN}

For generic values of $\eta$ (at fixed $\rho$), the Lagrangian ${\cal L}_{\rm toy}$ is not invariant under the fermionic chiral transformations $\tilde\chi_{L} \times \tilde\chi_{R}$ of eqs.~(\ref{GTWTL})-(\ref{GTWTR}), because of the presence of the (chiral breaking) operators ${\cal L}_{Wil}$ and ${\cal L}_{Yuk}$. Nevertheless, invariance under $\tilde\chi_{L} \times \tilde\chi_{R}$ can be recovered (up to O($b^{2}$) terms) by enforcing the conservation of the corresponding currents. This can be seen to occur~\cite{Frezzotti:2014wja,Frezzotti:2013raa} at a ``critical'' value of the Yukawa coupling, $\eta_{cr}$, where the $\tilde\chi_{L} \times \tilde\chi_{R}$ rotations of the Yukawa and Wilson-like operators, that separately break current conservation, ``compensate'' each other. 

The situation here is very similar to what happens in WLQCD where fermionic chiral symmetry is recovered (up to O($a$) cutoff effects) by tuning the bare quark mass to a critical value, $m_{cr}$, at which the chiral rotations of the Wilson and mass operators ``compensate'' each other, yielding the conservation of the chiral currents~\cite{WilsonLQCD,Bochicchio:1985xa}.\  As remarked above, a key difference between the model~(\ref{SULL}) and WLQCD is, however, that in WLQCD no symmetry exists that can prevent the appearance of a linearly divergent quantum correction to the fermion mass, unlike what happens in the case of the model~(\ref{SULL}). 

Naturally, as was done in ref.~\cite{Capitani:2019syo}, the determination of the critical value of the Yukawa coupling needs to be performed in a non-perturbative way, similarly to what is currently done in WLQCD in the calculation of the critical mass. Nevertheless, in order to get a feeling of what this criticality condition implies, it is interesting to see what happens at the lowest loop order. 

\begin{figure}[htbp]    
\centerline{\includegraphics[scale=0.35,angle=0]{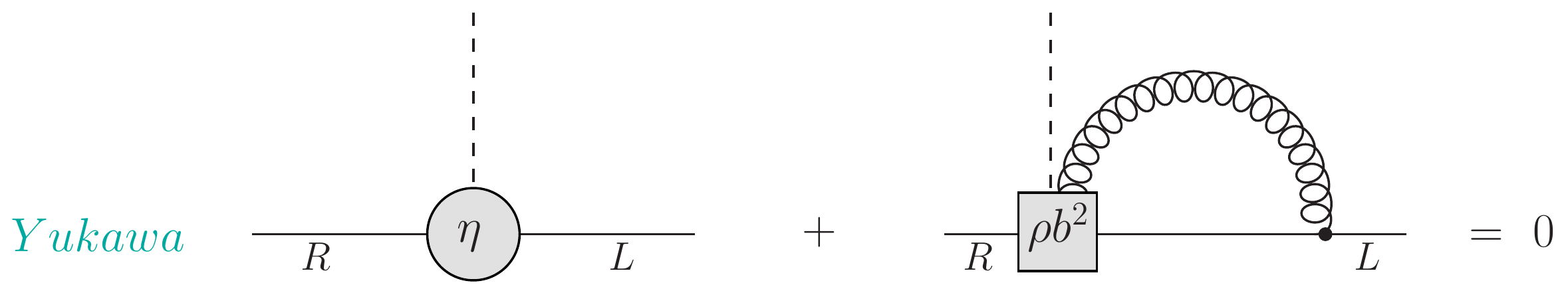}}
\caption{\small{The lowest loop order mechanism behind the cancellation of the Yukawa interaction in the Wigner phase occurring at $\eta=\eta_{cr}$. Grey disc and box stand for the insertion of the Yukawa and Wilson-like Lagrangian vertices, respectively. The solid line represents the propagation of a fermion, the curly line of a gluon and the dotted line of a scalar. {\it R} and {\it L} under the fermion line denote chirality. There is a similar condition with $R$ and $L$ exchanged.}}
\label{fig:fig1}
\end{figure}

$\bullet$ In the Wigner phase (where $\langle |\Phi|^2\rangle=0$) the condition enforcing the conservation of the $\tilde\chi_{L} \times \tilde\chi_{R}$ currents (derived  in~\cite{Frezzotti:2014wja} along the lines of the strategy worked out in~\cite{Bochicchio:1985xa} and summarized in Appendix~\ref{sec:APPA}) gives at 1-loop 
\beq
\eta_{cr}^{(\rm 1-loop)}(\alpha_s,\rho)= \rho \,\alpha_s N_c \eta_1\, ,
\label{ECR}
\eeq
where $\eta_1$ is a computable coefficient. The lowest loop order mechanism underlying this cancellation is sketched in fig.~\ref{fig:fig1}. At the NP level the critical condition implies the vanishing of the effective Yukawa coupling in the QEL of the theory.

\begin{figure}[htbp]    
\centerline{\includegraphics[scale=0.35,angle=0]{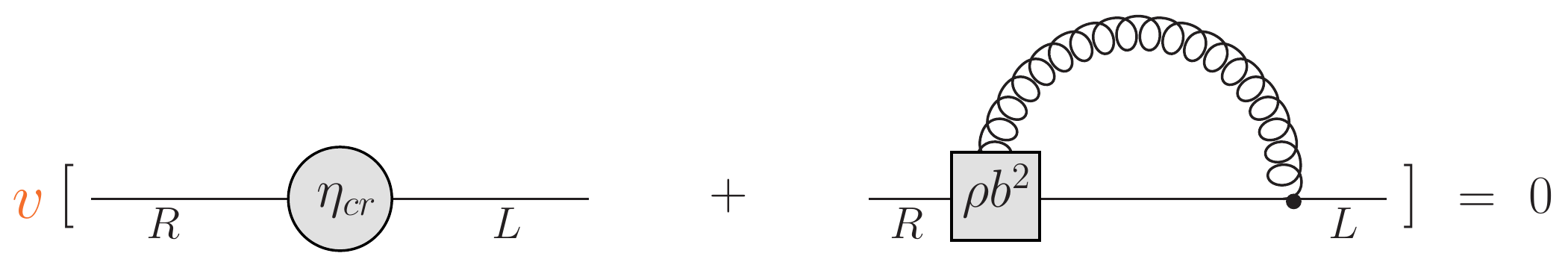}}
\caption{\small{The lowest loop order mechanism behind the cancellation of the Higgs-like mass of the quark in the NG phase, occurring at $\eta=\eta_{cr}$. There is a similar condition with $R$ and $L$ exchanged.}}
\label{fig:fig2}
\end{figure}

$\bullet$ In the NG phase (where $\langle |\Phi|^2\rangle\neq 0$ and at tree-level equal to $v^2=|\mu_0^2|/\lambda_0$), the $\eta$ criticality condition leads at the lowest loop order to the vanishing of the sum of diagrams shown in fig.~\ref{fig:fig2}. The figure is obtained from fig.~\ref{fig:fig1} by setting the scalar field equal to its vev. We thus see that, remarkably, enforcing $\tilde \chi_L\times \tilde \chi_R$ invariance implies that in the QEL of the critical theory precisely the Higgs-like mass of the fermion gets cancelled out.

We end this section with a few important remarks. 

1) The possibility of recovering $\tilde\chi_{L} \times \tilde\chi_{R}$ invariance rests on the crucial fact that the quadratic divergency of the loop integral in fig.~\ref{fig:fig1}, as well as in fig.~\ref{fig:fig2}, is exactly compensated by the $b^2$ factor coming from the insertion of the Wilson-like box, leaving behind a finite result.\ This balance of UV and IR effects (which has its ultimate basis on dimensional grounds) should not come as a surprise when irrelevant operators are present in the Lagrangian. Indeed, compensations of this kind also occur in WLQCD and are, for instance, at the origin of the fact that the (partially) conserved axial and vector flavour currents suffer a finite renormalization~\footnote{Actually in WLQCD in the vector case one can construct a current that, being exactly conserved at any value of the lattice spacing, does not need any renormalization.}. In the following we shall encounter other instances of this sort of ``UV vs.\ IR compensation''. 

2) We see thus that (owing to the cancellation pictorially illustrated at the lowest loop order in fig.~\ref{fig:fig2}) in the present approach $\Phi$ is not going to play the role the Higgs field does in the SM, rather its presence should be viewed as an effective, simple way to model a $\chi_L\times\chi_R$ invariant UV completion of ${\cal L}_{\rm toy}$. 

3) According to the 't Hooft criterion~\cite{THOOFT}, tuning $\eta$ at its critical value is a costless and natural step as it leads to an enlargement of the set of symmetries of the Lagrangian.

\subsection{The QEL of the critical theory in the Wigner phase}
\label{sec:QELW}

In this section we provide the form that the QEL takes in the Wigner phase at the critical value of the Yukawa coupling. As we said, enforcing invariance under $\tilde\chi_L\times\tilde\chi_R$ transformations means that $\tilde\chi$-violating operators should be absent from the expression of the QEL of the critical theory. The $d\leq 4$ piece of the QEL of the critical model in the Wigner phase will then have the simple form
\beqn
\hspace{-1.6cm}&&{\Gamma}_{4\,cr}^{Wig}=\frac{1}{4} F^A\!\cdot\! F^A+
\big{(} \bar q_L\,\Dslash^{A} q_L+\bar q_R\,\Dslash^{A} q_R\big{)} +\frac{1}{2}{\tr}\big{[}\partial_\mu\Phi^\dagger\partial_\mu\Phi\big{]}+{\cal V}(\Phi)\, ,
\label{L4WIG1}
\eeqn 
from which we see that scalars are completely decoupled from fermions. We should observe that in the Wigner phase the critical theory is kind of ``unstable''. In fact, strictly speaking, at the critical point no seed exists that can trigger the phenomenon of the spontaneous breaking of the (restored) $\tilde\chi_L\times\tilde\chi_R$ chiral symmetry. However, any chiral breaking disturbance, no matter how small, would make spontaneous breaking to take place~\cite{WEIN} with the vacuum choosing its orientation.

\subsection{NP mass generation in the NG phase}
\label{sec:NPMG}

The situation one encounters in the NG phase of the critical theory is totally different. We shall see that, despite the fact that in the NG phase the Higgs-like mass of the fermion gets cancelled (see fig.~\ref{fig:fig2}), a non-vanishing NP fermion mass-like term emerges. 

Because of the subtle UV vs.\ IR interplay that we have seen is at work in the model, in order to get some understanding of how a non-vanishing NP fermion mass can emerge, it is crucial to start the discussion by determining the structure of the possible operators of NP origin, formally of O($b^2$), arising in the regularized theory. The analysis can be properly done by making use of the Symanzik expansion technique~\cite{Symanzik:1983gh,Symanzik:1983ghh}. 

A study of this kind was carried out in ref.~\cite{Frezzotti:2014wja} where it was shown that, as a consequence of the occurrence of the spontaneous breaking of the (restored) $\tilde\chi_L\times\tilde\chi_R$ symmetry (which is in turn triggered by residual O($b^2 v$) chiral breaking terms surviving at $\eta_{cr}$ in the NG phase), NP-ly generated Symanzik operators of O($b^2\Lambda_s$) emerge. These NP effects can be described and included in the theory if, following refs.~\cite{Symanzik:1983gh,Symanzik:1983ghh}, we allow ourself to work with the ``augmented'' Lagrangian 
\beqn
\hspace{-1.6cm}&&{\cal L}_{\rm toy} \to {\cal L}_{\rm toy} + \Delta{\cal L}_{NP} \, , \qquad \Delta{\cal L}_{NP}=\gamma_{\bar q q} (\alpha_s) O_{6,\bar q q} + \gamma_{AA} (\alpha_s)O_{6,AA} \, ,\label{LAUG}\\
\hspace{-1.6cm}&&O_{6,\bar q q} = b^2 \Lambda_s |\Phi|\, \big{(}\bar q_L\,\, \Dslash^{A} q_L\!+\!\bar q_R\,\, \Dslash^{A} q_R\big{)} \, , \qquad
O_{6,AA} = b^2\Lambda_s|\Phi| \, F^A\!\cdot\!F^A \, ,\label{SOP}
\eeqn
where the coefficients $\gamma_{\bar q q}(\alpha_s)$ and $\gamma_{FF}(\alpha_s)$ are functions of the gauge coupling~\footnote{As usual, we define $\alpha_s=g^2_s/4\pi$.}. More in detail, extending the arguments given in ref.~\cite{Frezzotti:2014wja}, one can show that at lowest order we find that both $\gamma_{\bar q q}(\alpha_s)$ and $\gamma_{AA}(\alpha_s)$ are ${\mbox{O}}(\alpha_s)$, if a $d=6$ Wilson-like term, as in eq.~(\ref{LWIL}), appears in the fundamental Lagrangian~(\ref{SULL}). Complementary arguments supporting  the emergence of such NP operators are developed in~Appendix~B of~(II). 

The structure of the operators $O_{6,\bar q q}$ and $O_{6,AA}$ is completely dictated by symmetries (in particular $\chi_L\times\chi_R$) and dimensional considerations. The presence of the $\Lambda_s$ factor signals their NP origin. According to the rules of the Symanzik analysis~\cite{Symanzik:1983gh,Symanzik:1983ghh}, making reference to ${\cal L}_{\rm toy} +\Delta{\cal L}_{NP}$ should be seen as a bookkeeping of the NP operator insertions occurring in the correlators of the fundamental Lagrangian. Extending in this way the allowed set of diagrams is necessary to fully describe all the NP features of the theory in the NG phase. In fact, though formally of O($b^2$), the insertions of $\Delta{\cal L}_{NP}$ cannot be ignored, because, as we have repeatedly said, explicit $b^2$ factors can be compensated by UV power divergencies in loop integrals, eventually leading to finite NP contributions to correlators. 

\begin{figure}[htbp]    
\centerline{\includegraphics[scale=0.4,angle=0]{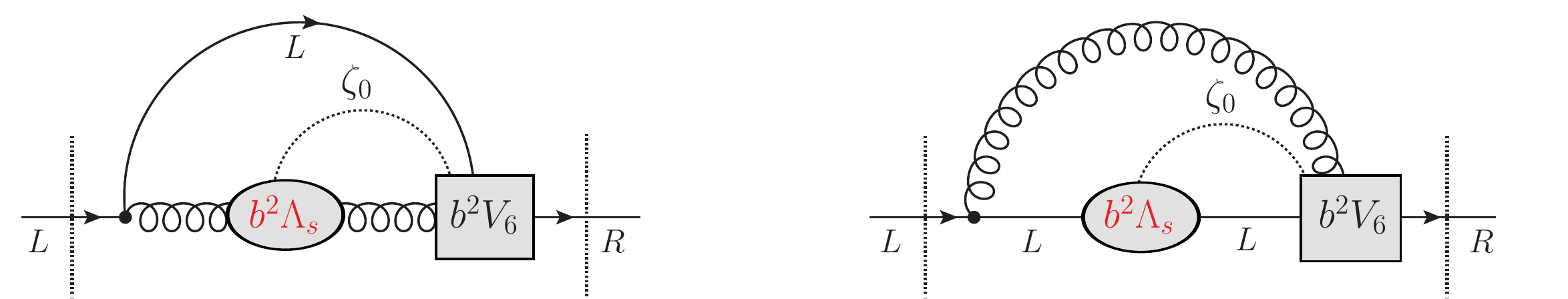}} 
\caption{\small{Two typical lowest loop order NP quark self-energy diagrams. The blobs represent  insertions of the operators $O_{6,FF}$ and $O_{6,\bar q q}$, respectively, and the dotted line the propagation of the singlet scalar, $\zeta_0$, defined in eq.~(\ref{PHPH}) below. The vertical lines mean amputation. The rest of the notations is as in fig.~\ref{fig:fig1}.}}
 \label{fig:fig3}
\end{figure}

The remarkable fact about the ``augmented'' Lagrangian~(\ref{LAUG}) is that it generates among others, new diagrams of NP origin (surviving in the $b\to 0$ limit) that contribute to the quark self-energy. Two instances of such diagrams are drawn in fig.~\ref{fig:fig3}.\ An explicit calculation of the amputated zero momentum diagram shown in the right panel of fig.~\ref{fig:fig3} gives to the effective fermion mass the finite contribution~\footnote{The vertical dotted line means amputation of the external leg. We have nevertheless left explicit the ingoing and outgoing lines to help the reader recognizing the particles one is referring to.} 
{\begin{eqnarray}
\hspace{-1.2cm}&&m_q \propto \alpha_s^2\, \tr \!\!\int^{1/b}\!\frac{d^4 k}{k^2}\frac{\gamma_\mu k_\mu}{k^2} \!\int^{1/b}\!\!\!\!\frac{d^4 \ell}{\ell^2+m^2_{\zeta_0}} \frac{\gamma_\nu(k+\ell)_\nu}{(k+\ell)^2} \, b^2\gamma_\rho (k+\ell)_\rho \,b^2 {\Lambda_{s}}\gamma_\lambda (2k+\ell)_\lambda \!\sim  {\alpha_s^2 \Lambda_{s}} \label{MQ}\, ,
\end{eqnarray}
where one factor of $\alpha_s$ comes from the gluon connecting a standard QCD-like vertex with the Wilson-like box and another from the insertion of the $O_{6,\bar q q}$ operator in eq.~(\ref{SOP}) with the coefficient $\gamma_{\bar q q}$ taken at its lowest order. In the integrand of eq.~(\ref{MQ}) $m^2_{\zeta_0}=2|\mu^2_0|$ is the (square) mass of the singlet $\zeta_0$ which is standardly defined by the polar decomposition 
\beqn
&&\Phi = RU\, , \quad R=v+\zeta_0\, , \label{PHPH} \\
&&U=\exp[i\vec\tau\,\vec \zeta/c\Lambda_s]\, ,\label{PHPH2}
\eeqn
with the (arbitrary) scale in the exponential conveniently chosen with an eye to the QEL in eq.~(\ref{GW}) below, so as to have the effective NG fields $\zeta^i, i=1,2,3$ canonically normalized. Obviously $U$ can only be defined in the NG phase of the theory where $\langle |\Phi|^2\rangle\! \neq \!0$.

We end the section with a few remarks.

1) First of all, since fermion masses are running quantities, we must specify which scale the mass computed in eq.~(\ref{MQ}) refers to. We note that $m_q$ emerges from diagrams that are completely dominated by the UV behaviour of the integrated loop momenta. Thus we conclude that the self-energy diagrams of fig.~\ref{fig:fig3} provide the fermion mass at asymptotically large scale, or better, if the theory unifies, at the GUT scale. We will make use of this important observation in the phenomenological considerations we will present in sect.~5 of ref.~(II).

2) Secondly, we need to explain what we mean by ``$\zeta_0$-propagator'', given the fact that the critical theory is defined precisely at the value of $k_b$ where the kinetic term of $\Phi$ is canceled against the similar operator with which the Wilson-like terms mix. This point is mentioned in sect.~\ref{sec:ZZP} and fully clarified in Appendix~\ref{sec:APPC}. 

3) For phenomenological considerations (see sect.~5 in~(II)) it is important to decide whether we should read the coupling constant dependence in the last line of eq.~(\ref{MQ}) as $\alpha_s^2$ (as indicated), or as $\alpha_s\,g_s^2$, or rather as $(g_s^2)^2$, because these choices differ by non-irrelevant numerical factors. This question will be taken up in sect.~\ref{sec:NPM} and discussed in detail in Appendix~B of ref.~(II) where a criterion to determine the form of the ``optimal'' gauge coupling dependence of the $\gamma$-coefficients (i.e.\ the choice that correctly absorbs the $1/4\pi$ factors arising in loops) is proposed/provided. 

4) To simplify the line of reasoning leading to the key eq.~(\ref{MQ}) we have provisionally ignored the fact that in the operators~(\ref{SOP}) it is $|\Phi|$ (i.e.\ the modulus of $\Phi$) that appears and not $\Phi$ itself. This implies that in the diagrams of fig.~\ref{fig:fig3} after the $\zeta_0$ contraction a ``dandling'' $U$ factor is left out exiting the Wilson-like box (recall eq.~(\ref{PHPH})). We did not show this field dependence in fig.~\ref{fig:fig3}, because for the calculation of the mass one can set the external field $U$ equal to the identity. However, the complete $U$ dependence of these diagrams needs to be reinserted in order to be able to generate a fully $\chi_L\times\chi_R$ invariant fermion mass operator, like the first term in the second line of eq.~(\ref{GW}) below.

\subsection{The QEL of the critical theory in the NG phase}
\label{sec:QNGC}

The existence of a NP fermion mass term, the occurrence of which in the NG phase of the model~(\ref{SULL}) was also successfully checked in ref.~\cite{Capitani:2019syo} by explicit lattice simulations, can be incorporated in the QEL that describes the physics of the theory (denoted by $\Gamma^{NG}_{cr}$ in the following) with the help of the non-analytic field $U$ introduced in eq.~(\ref{PHPH}). Owing to the fact that $U$ has the same transformation properties under $\chi_L\!\times\!\chi_R$ as $\Phi$ does, new operators (with respect to those appearing in the critical QEL of the Wigner phase, eq.~(\ref{L4WIG1})) invariant under $\chi_L\!\times\!\chi_R$ can be constructed, leading for the $d\leq 4$ piece of $\Gamma^{NG}_{cr}$ to the expression 
\beqn
\hspace{-.9cm}&& \Gamma_{4\,cr}^{NG}\! =\! 
\frac{1}{4}F^A\!\cdot\! F^A\!+\! \big{(}\bar q_L \,\Dslash^{A} q_L\!+\!\bar q_R\, \Dslash^{A} q_R
\big{)} \!+\nn\\
\hspace{-.9cm}&&\qquad+c_q \Lambda_s \big{(}\bar q_L U q_R\! +\!\bar q_R U^\dagger q_L\big{)} \!+\!\frac{c^2\Lambda_s^2}{2} \tr[\partial_\mu U^\dagger\partial_\mu U]\!+\ldots \label{GW}
\eeqn
with $c_q$ and $c$ functions of the gauge coupling. The form of $\Gamma_{4\,cr}^{NG}$ is purely geometrical. It is constrained by dimensional and symmetry arguments and the observation that at $\eta=\eta_{cr}$ the $\tilde\chi_L\times\tilde\chi_R$ breaking Yukawa term should not be present. The expression~(\ref{GW}) is obtained by including all the operators of dimension $d\leq 4$, invariant under $\chi_L\times\chi_R$ that can be constructed in terms of $q$, $\bar q$, $A_\mu$ and $U$. 

Dots in~(\ref{GW}) are there to recall us that actually there exist other operators invariant under $\chi_L\!\times\!\chi_R$ that we have not indicated. At $d=4$ we have the scalar kinetic term $\frac{1}{2}{\tr}\big{[}\partial_\mu\Phi^\dagger\partial_\mu\Phi\big{]}$ as well as the operator $\Lambda_s R\,\tr[\partial_\mu U^\dagger\partial_\mu U]$ and the scalar potential ${\cal V}(\Phi)$. However, as we shall argue in the next section (see Appendix~\ref{sec:APPB}), in the presence of weak interactions, the restoration of $\tilde\chi_L\times \tilde\chi_R$ makes the first two operators to disappear from the QEL. At the same time, at the critical point the effective singlet field $R=v+\zeta_0$ (eq.~(\ref{PHPH})) becomes infinitely massive and decouples. 

Naturally, the third term in the r.h.s.\ of eq.~(\ref{GW}) (which is not invariant under $\tilde\chi_L\times\tilde\chi_R$ transformations) is of special interest because, upon expanding $U =\!1\!\!1\!+ i \vec \tau\,\vec\zeta/c\Lambda_s+\ldots$ (see eq.~(\ref{PHPH2})), it gives rise to a mass for the fermion plus a wealth of NG boson non-linear interactions. 

We observe that, despite the fact that the tuning of the Yukawa coupling enforces the conservation of the chiral $\tilde\chi_L\times\tilde\chi_R$ currents, we have discovered that peculiar mass-like terms are dynamically generated which represent NP obstructions to the exact realization of chiral symmetry~\footnote{Actually also in QCD mass terms break chiral symmetry. The difference is that, while in QCD they can be put to zero by hand, here masses are NP effects that cannot be eliminated, as they arise as soon as $\rho\neq 0$ no matter how small $|\rho|$ might be.}. A suggestive way to describe the nature of this dynamically generated mass term is to say that the latter appears as a sort of ``NP anomaly'' preventing the full recovery of the chiral $\tilde\chi_{L} \times \tilde\chi_{R}$ symmetry.

\section{Introducing weak and Tera-interactions}
\label{sec:WMASS}

To proceed to the construction of a realistic model for elementary particles physics the next step is to introduce weak interactions. At the same time, as mentioned in the Introduction, it is also necessary to extend the model by incorporating a super-strongly interacting sector, so that the whole theory will have an RGI scale, $\Lambda_{T}$, much larger than $\Lambda_{QCD}$ and, to match phenomenology, of the order of a few TeVs. Only in this way eqs.~(\ref{EQ1}) and~(\ref{EQ2}) can possibly yield the correct order of magnitude of the top quark and $W$ boson mass, respectively. 

The desired extension of the model is obtained by doubling the structure of fermions to encompass Tera-particles ($Q\!=$\,Tera-quarks and $G\!=$\,Tera-gluons) and introducing weak interactions by gauging the exact $\chi_L$ symmetry. The resulting Lagrangian will have the expression 
{\beqn
\hspace{-.7cm}&&{\cal L}(q,Q;A,G,W;\Phi)\!= \nn\\
\hspace{-.7cm}&&\qquad={\cal L}_{K}(q,Q;A,G,W;\Phi)\!+\!{\cal V}(\Phi)\!+\!{\cal L}_{Yuk}(q,Q;\Phi)\!+\!{\cal L}_{Wil}(q,Q;A,G,W;\Phi) \label{SULLWQ}\\
\hspace{-.7cm}&&\,\,\bullet\,\,{\cal L}_{K}(q,Q;A,G,W;\Phi)= \frac{1}{4}\Big{(}F^A\cdot F^A+F^G\cdot F^G+F^W\cdot F^W\Big{)}+\label{LKINWQ}\\
\hspace{-.7cm}&&\,\,\quad+\big{(}\bar q_L\,/\!\!\!\!{\cal D}^{AW} q_L\!+\!\bar q_R\,/\!\!\!\!{\cal D}^{A} q_R\big{)}\!+\!\big{(}\bar Q_L\,/\!\!\!\!{\cal D}^{AGW} Q_L\!+\!\bar Q_R\,/\!\!\!\!{\cal D}^{AG} Q_R\big{)}\!+\!\frac{k_b}{2}{\tr}\big{[}({\cal D}\,^{W}_\mu \Phi)^\dagger{\cal D}^W_\mu\Phi\big{]}\nn\\
\hspace{-.7cm}&&\,\,\bullet\,\,{\cal V}(\Phi)= \frac{\mu_0^2}{2}k_b{\tr}\big{[}\Phi^\dagger\Phi\big{]}+\frac{\lambda_0}{4}\big{(}k_b{\tr}\big{[}\Phi^\dagger\Phi\big{]}\big{)}^2\label{VWQ}\\
\hspace{-.7cm}&&\,\,\bullet\,\,{\cal L}_{Yuk}(q,Q;\Phi)=\eta_q\,\big{(} \bar q_L\Phi\, q_R+\bar q_R \Phi^\dagger q_L\big{)} + \eta_Q\,\big{(} \bar Q_L\Phi\, Q_R+\bar Q_R \Phi^\dagger Q_L\big{)}\label{LYUKWQ} \\
\hspace{-.7cm}&&\,\,\bullet\,\,{\cal L}_{Wil}(q,Q;A,G,W;\Phi)= \frac{b^2}{2}{\rho_q}\,\big{(}\bar q_L{\overleftarrow {\cal D}}\,^{AW}_\mu\Phi {\cal D}^A_\mu q_R+\bar q_R \overleftarrow{\cal D}\,^A_\mu \Phi^\dagger {\cal D}^{AW}_\mu q_L\big{)}+\nn\\
\hspace{-.7cm}&&\,\,\quad+\frac{b^2}{2}{\rho_Q}\,\big{(}\bar Q_L{\overleftarrow {\cal D}}\,^{AGW}_\mu\Phi {\cal D}^{AG}_\mu Q_R+\bar Q_R \overleftarrow{\cal D}\,^{AG}_\mu \Phi^\dagger {\cal D}^{AGW}_\mu Q_L\big{)}\label{LWILWQ}
\eeqn
with obvious notations for the covariant derivatives. As for the symmetries of the Lagrangian~(\ref{SULLWQ}), the previous form of the $\chi_L\times\chi_R$ (eq.~(\ref{CHIL})) and $\tilde\chi_L\times\tilde\chi_R$ (eqs.~(\ref{GTWTL})-(\ref{GTWTR})) transformations, besides the obvious extension necessary to let them act also on Tera-fermions, need to be modified in the presence of $W$ bosons to insure invariance of the fermionic kinetic terms under $\tilde\chi_L\times\tilde\chi_R$. The analog of the set of eqs.~(\ref{CHIL}), (\ref{GTWTL}) and~(\ref{GTWTR}) then reads ($\Omega_{L/R}\!\in$SU(2))
\begin{eqnarray}
\hspace{-.5cm}&&\chi_L\times \chi_R =  [\tilde\chi_L\times (\Phi\to\Omega_L\Phi)]\times [\tilde\chi_R\times (\Phi\to\Phi\,\Omega_R^\dagger)] \, , 
\label{CHILN}\\
\hspace{-1.4cm}&&\tilde\chi_L : \left \{\begin{array}{l}     
q_L\rightarrow\Omega_L q_L  \qquad\qquad
\quad\bar q_L\rightarrow \bar q_L\Omega_L^\dagger \\
Q_L\rightarrow\Omega_L Q_L  \qquad\qquad
\,\bar Q_L\rightarrow \bar Q_L\Omega_L^\dagger \\
W_\mu \rightarrow\Omega_L W_\mu \Omega_L^\dagger\\
\end{array}\right . \label{GTWTQ}\\ 
\hspace{-2.cm}&&\tilde\chi_R : \left \{\begin{array}{l}     
q_R\rightarrow\Omega_R q_R  \qquad\qquad
\quad\bar q_R\rightarrow \bar q_R\Omega_R^\dagger \\
Q_R\rightarrow\Omega_R Q_R \qquad\qquad
\,\bar Q_R\rightarrow \bar Q_R\Omega_R^\dagger
\end{array}\right . \label{GTCTQ}
\end{eqnarray}

\subsection{The critical theory}
\label{sec:TCT}

Besides the Yukawa (eq.~(\ref{LYUKWQ})) and the Wilson-like (eq.~(\ref{LWILWQ})) operators, now also the kinetic term of the scalar field (last term in eq.~(\ref{LKINWQ})) breaks $\tilde\chi_L\times \tilde\chi_R$ and mixes with ${\cal L}_{Yuk}$ and ${\cal L}_{Wil}$.\ Thus to get the critical theory (invariant under $\tilde\chi_L\times \tilde\chi_R$), on top of the Yukawa couplings $\eta_q$ and $\eta_Q$, a further parameter, $k_{b}$, has been introduced which needs to be appropriately tuned. For convenience the coefficient $k_b$ is let to appear also in the expression of the scalar potential~(\ref{VWQ}), because with this choice the bare (vev)$^2$ (in the NG phase) and the quartic coupling of the canonically normalized scalar field will keep their standard definitions, i.e.\ $v^2 = |\mu^2_0|/\lambda_0$ and $\lambda_0$.

The tuning conditions determining $\eta_{q\,cr}$, $\eta_{Q\,cr}$ and $k_{b\,cr}$ come from enforcing in the Wigner phase the conservation of the $\tilde\chi_L\times\tilde\chi_R$ currents~\cite{Bochicchio:1985xa}.\ In Appendix~A for completeness we sketch the procedure that, extending the strategy proposed in~\cite{Frezzotti:2014wja}, leads to the tuning equations for $\eta_q$, $\eta_Q$ and $k_b$. The conditions determining the critical theory physically correspond to have no scalar kinetic term in the QEL of the theory and, similarly to what we saw happening in sect.~\ref{sec:STM} in the case of the model~(\ref{SULL}), vanishing effective Yukawa interactions. 

Naturally, as was done in ref.~\cite{Capitani:2019syo} for the Yukawa coupling, the computation of the critical values of $\eta_q$, $\eta_Q$ and $k_b$ needs to be performed in a non-perturbative way, like it is currently done in WLQCD in the determination of the critical mass. Nevertheless, as we did in sect.~\ref{sec:CHIN}, in order to get a feeling of what these criticality conditions mean, it is interesting to examine what happens at the lowest loop order.                  

\subsection{The critical conditions in the Wigner phase}
\label{sec:CCWP}

Solving, with steps similar to those leading to eq.~(\ref{ECR}), the tuning eqs.~(\ref{CREQSQ1})-(\ref{CREQSQ5}) one finds for $\eta_{q\,cr}$, $\eta_{Q\,cr}$ and $k_{b\,cr}$ at 1-loop in the Wigner phase the parametric expressions
\beqn
\hspace{-.4cm}&&\eta_{q\,cr}^{(\rm 1-loop)}=\rho_q \,  \alpha_s N_c \eta_{q}^{(1)} \, , \qquad\qquad\qquad \eta_{Q\,cr}^{(\rm 1-loop)}=\rho_Q \,\alpha_T  N_T\eta_{Q}^{(1)}  \, ,\label{ETAT1}\\
\hspace{-.4cm}&&\hspace{2.cm}k_{b\,cr}^{(\rm 1-loop)}= \rho_q^2 N_c k_{b\,q}^{(1)} + \rho_Q^2 N_c  N_T k_{b\,Q}^{(1)} \, ,\label{KCR1}
\eeqn
with $\eta_{q}^{(1)}$, $\eta_{Q}^{(1)}$, $k_{b\,q}^{(1)}$ and $k_{b\,Q}^{(1)}$ computable coefficients and SU($N_T$) the Tera-strong gauge group.
 
In figs.~\ref{fig:fig4a} and~\ref{fig:fig4b} we display at the lowest loop order the mechanism leading to the cancellation of the effective quark and Tera-quark Yukawa terms in the QEL of the critical theory. The figures show the mixing between the quark (Tera-quark) Yukawa operator (grey disks) and the quark (Tera-quark) Wilson-like operator (grey box) at 1-loop.

In fig.~\ref{fig:fig4c} we report the leading 1-loop diagrams yielding the cancellation between the scalar kinetic term (grey disk) and the contribution from the sum of the quark and Tera-quark Wilson-like operators (grey boxes).\ The empty boxes represent the insertion of Wilson-like vertices from the Lagrangian necessary to close the fermion loops. As before, the loop diagrams in the figs.~\ref{fig:fig4a}, \ref{fig:fig4b} and~\ref{fig:fig4c} yield finite results because the loop UV power divergencies are exactly compensated by the $b^2$ factors coming from the insertion of Wilson-like vertices.

\begin{figure}[htbp]
\centerline{\includegraphics[scale=0.3,angle=0]{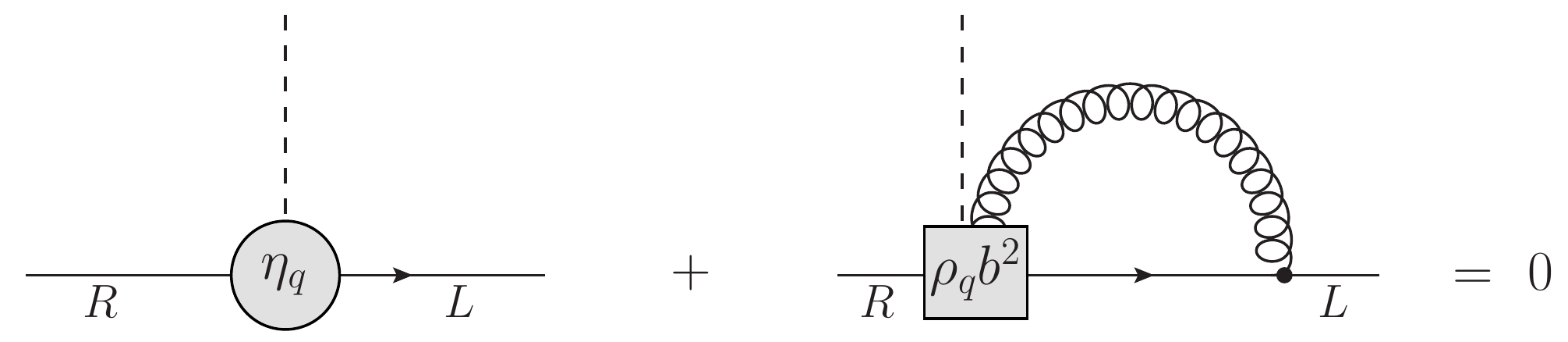}}
\caption{\small{The cancellation of the quark Yukawa vertex implied by the tuning conditions determining $\eta_{q\,cr}$ at the lowest loop order in the Wigner phase. The grey box, labelled by $\rho_q$, represents the quark Wilson-like vertex. The rest of the notations is as in fig.~\ref{fig:fig1}.}}
\label{fig:fig4a}
\end{figure}
\begin{figure}[htbp]
\centerline{\includegraphics[scale=0.3,angle=0]{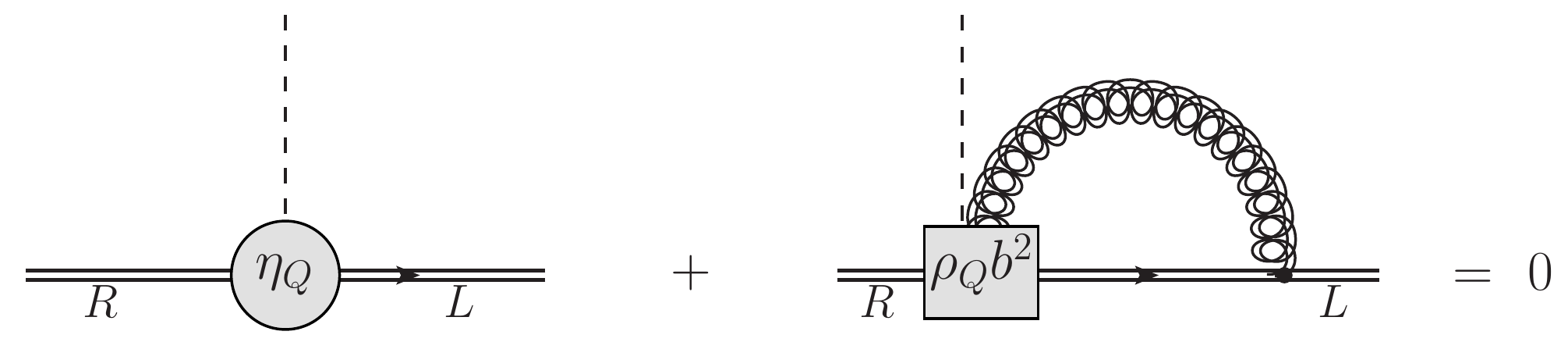}}
\caption{\small{The cancellation of the Tera-quark Yukawa vertex implied by the tuning conditions determining $\eta_{Q\,cr}$ at the lowest loop order in the Wigner phase. Double lines represent Tera-particles. The grey box, labelled by $\rho_Q$, represents the Tera-quark Wilson-like vertex. The rest of the notations is as in fig.~\ref{fig:fig4a}.}}
\label{fig:fig4b}
\end{figure}
\begin{figure}[htbp]
\centerline{\includegraphics[scale=0.35,angle=0]{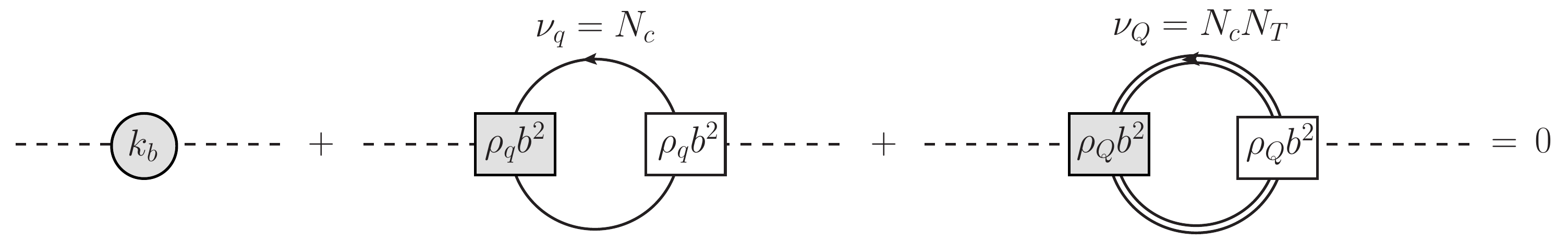}}
\caption{\small{The cancellation of the scalar kinetic term implied by the tuning condition determining $k_{b\,cr}$ at the lowest loop order in the Wigner phase. The integers $\nu_q$ and $\nu_Q$ are the multiplicities of quarks and Tera-quarks running in the loops with $N_c$ and $N_T$ the number of colours and Tera-colours, respectively. The grey disc, labelled by $k_b$, represents the insertion of the scalar kinetic term. The empty boxes represent the insertion of Wilson-like vertices from the Lagrangian. The rest of the notations is as in figs.~\ref{fig:fig4a} and~\ref{fig:fig4b}.
}}
\label{fig:fig4c}
\end{figure}

\subsection{The critical conditions in the Nambu--Goldstone phase}
\label{sec:CCNGP}

In the NG phase the criticality conditions for $\eta_q$, $\eta_Q$ and $k_b$ imply  the cancellations shown in the figs.~\ref{fig:fig5a}, \ref{fig:fig5b} and~\ref{fig:fig5c}, respectively. Figs.~\ref{fig:fig5a} and~\ref{fig:fig5b}, as in sect.~\ref{sec:STM}, are directly obtained from the figs.~\ref{fig:fig4a} and~\ref{fig:fig4b} by setting the scalar field equal to its vev. Fig.~\ref{fig:fig5c} follows from fig.~\ref{fig:fig4c} using SU(2)$_L$ gauge invariance and setting the scalar at its vev. 

Like we observed in sect.~\ref{sec:CHIN}, also here the key feature is that the NP enforcement of the $\tilde \chi_L\times \tilde \chi_R$ invariance implies that in the QEL of the critical theory precisely the Higgs-like masses of quarks, Tera-quarks and $W$'s are canceled out.
\begin{figure}[htbp]
\centerline{\includegraphics[scale=0.35,angle=0]{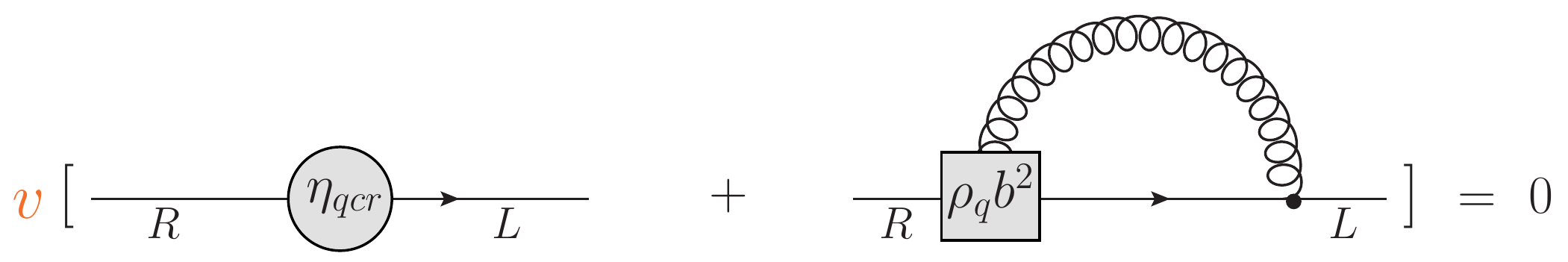}}
\caption{\small{The mechanism underlying the cancellation of the Higgs-like mass term of quarks at the lowest loop order occurring in the NG phase of the critical theory. Notations are as in fig.~\ref{fig:fig4a}.}}\label{fig:fig5a}
\end{figure}
\begin{figure}[htbp]
\centerline{\includegraphics[scale=0.35,angle=0]{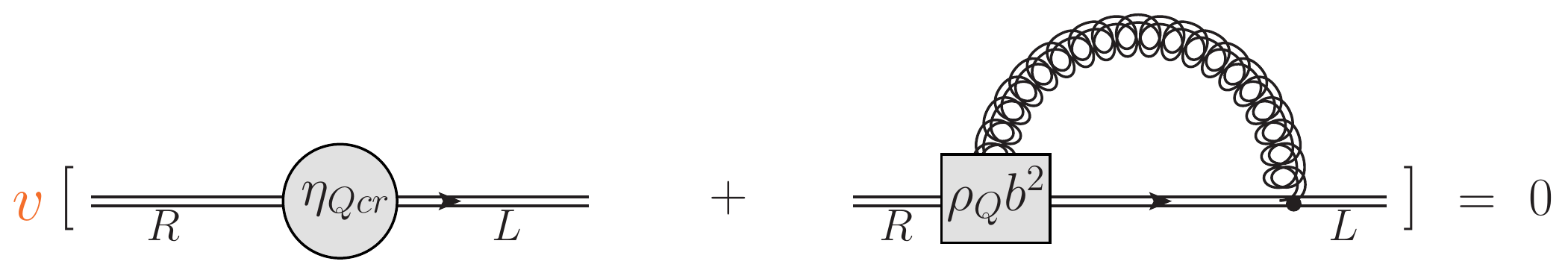}}
\caption{\small {The mechanism underlying the cancellation of the Higgs-like mass term of Tera-quarks at the lowest loop order occurring in the NG phase of the critical theory. Notations are as in fig.~\ref{fig:fig4b}.}} 
\label{fig:fig5b}
\end{figure}
\begin{figure}[htbp]
\centerline{\includegraphics[scale=0.35,angle=0]{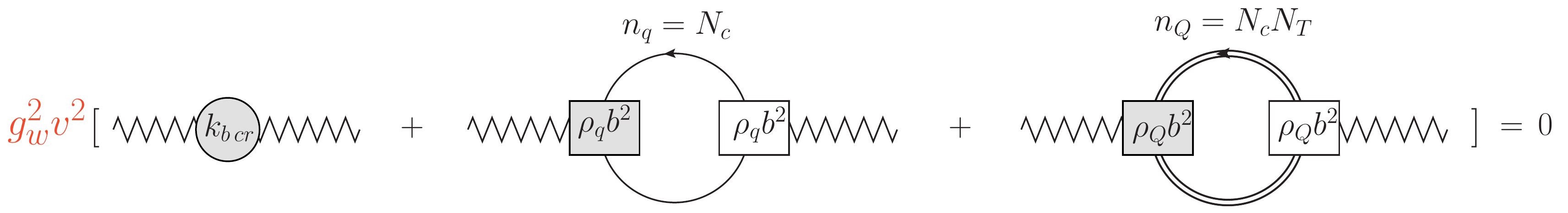}}
\caption{\small {The mechanism underlying the cancellation of the Higgs-like $W$ mass term at the lowest loop order occurring in the NG phase of the critical theory. Wiggly lines are $W$'s. The rest of the notations is as in figs.~\ref{fig:fig4a}, \ref{fig:fig4b} and~\ref{fig:fig4c}.}}
\label{fig:fig5c}
\end{figure}

\subsection{The NP emergence of elementary particles masses}
\label{sec:EEPM}

By extending to the present case the analysis that in sect.~\ref{sec:NPMG} has led us to identify the operators~(\ref{SOP}), one finds at O($b^2$) the following set of NP $\chi_L\times\chi_R$ invariant operators
\beqn
\hspace{-.6cm}&&\qquad O_{6,\bar Q Q}^T = b^2 \Lambda_T|\Phi|\, \big{(}\bar Q_ L\,\Dslash^{AGW} Q_L+\bar Q_R\,\Dslash^{AG} Q_R\big{)} \label{OSS} \\
\hspace{-.6cm}&&\qquad O_{6,GG} = b^2 \Lambda_T |\Phi| \,F^G\!\cdot\!F^G \label{OFFG}\\
\hspace{-.6cm}&&\qquad O_{6,AA} = b^2\Lambda_T |\Phi| \,F^A\!\cdot\!F^A \label{OFFA}
\eeqn
Although formally of O($b^2$), the effect of insertion of these operators in correlators cannot be ignored because of the kind of IR-UV compensation we have seen occurring when ``irrelevant'' operators are present in the fundamental Lagrangian. Taking into account the operators~(\ref{OSS})--(\ref{OFFA}) is necessary to fully describe the NP features of the theory related to the spontaneous breaking of the (restored) chiral symmetry. 

Similarly to the situation we discussed in sect.~\ref{sec:NPMG}, the occurrence  of the Symanzik NP operators~(\ref{OSS})--(\ref{OFFA}) can be properly taken into account by adding them to the fundamental Lagrangian, yielding the ``augmented'' Lagrangian~\cite{Symanzik:1983gh,Symanzik:1983ghh} 
\beqn
\hspace{-1.1cm}&&{\cal L}\to {\cal L}+ \Delta{\cal L}_{NP} 
\, ,\quad \Delta{\cal L}_{NP}=\gamma_{\bar Q Q}^T O_{6,\bar Q Q}^T + \gamma_{GG} O_{6,GG} +\gamma_{AA} O_{6,AA}  \, ,\label{AUGM}
\eeqn
where the coefficients $\gamma^T_{\bar Q Q}$, $\gamma_{GG}$ and $\gamma_{AA}$ are functions of the gauge couplings. In Appendix~B of~(II) we provide arguments supporting the existence of the NP operators~(\ref{OSS})--(\ref{OFFA}) and derive the lowest loop order behaviour of the $\gamma$-coefficients in eq.~(\ref{AUGM}) upon the gauge couplings~\footnote{Actually there exist other NP operators. An example is given by $\gamma_{WW}b^2\Lambda_T |\Phi| \,F^W\!\cdot\!F^W $, with $\gamma_{WW}={\mbox{O}}(g_w^2)$. The list above is limited to the operators that give rise to the leading diagrams contributing to the NP masses of fig.~\ref{fig:fig6}. }. As noted in sect.~\ref{sec:NPMG}, where we discussed the similar case of eq.~(\ref{LAUG}), eq.~(\ref{AUGM}) is purely formal, in the sense that, according to the rules of the Symanzik expansion~\cite{Symanzik:1983gh,Symanzik:1983ghh}, introducing ${\cal L} +\Delta{\cal L}_{NP}$ should be seen as a bookkeeping of the NP operator insertions occurring in the correlators of the fundamental Lagrangian, ${\cal L}$. 

At his point (see also our comment at the point 3) in the list at the end of sect.~\ref{sec:NPMG}) it is crucial to remark that, although unimportant for highlighting the parametric dependence of $\Delta{\cal L}_{NP}$ in eq.~(\ref{AUGM}) upon the gauge couplings, a key question for phenomenology is to decide whether we should use $g^2$ or $\alpha$ in order to best (``optimally'') parametrize the coefficients $\gamma$ with which the NP Symanzik operators~(\ref{OSS}), (\ref{OFFG}) and~(\ref{OFFA}) enter in ${\cal L} +\Delta{\cal L}_{NP}$. By ``optimal parametrization'' we mean choosing either $g^2$ or $\alpha=g^2/4\pi$ to absorb the factors $1/4\pi$ arising in loops. In Appendix~B of~ref.~(II) we show that, in the case of $d=6$ Wilson-like terms as those in eq.~(\ref{LWILWQ}), at lowest order, the optimal choice corresponds to the following behaviour
\beq
\gamma^T_{\bar Q Q}={\mbox{O}}(\alpha_T)\, ,\qquad \gamma_{GG} ={\mbox{O}}(g_T^2)\, ,\qquad \gamma_{AA}={\mbox{O}}(g^2_s)\, .
\label{GPROP}
\eeq 

\subsubsection{NP mass generation}
\label{sec:NPMGE}

The presence of new vertices leads among other contributions, to self-energy diagrams capable of generating NP masses not only for quarks, but also for Tera-quarks and weak bosons.\  At the lowest loop order one finds the typical amputated self-energy diagrams displayed in the four panels of fig.~\ref{fig:fig6}, where on a case by case basis the operators~(\ref{OSS}), (\ref{OFFG}) or~(\ref{OFFA}) are inserted together with the Wilson-like vertices necessary to close the loops. All these diagrams are finite owing to the by now familiar UV-IR compensation, and all of O($\Lambda_T$) times gauge coupling dependent coefficients.

\begin{figure}[htbp]
\centerline{\includegraphics[scale=0.7,angle=0]{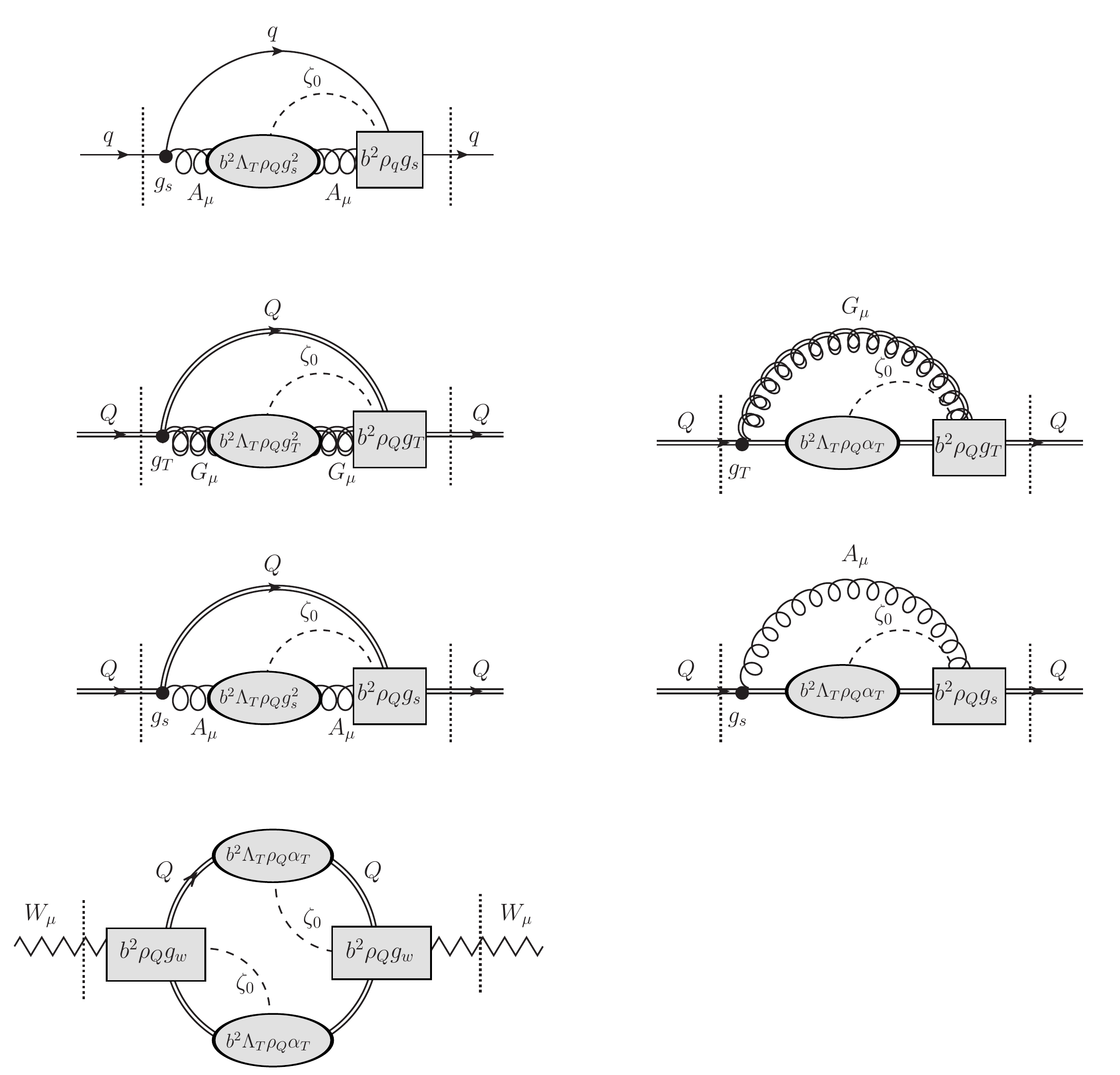}}
\caption{\small{Examples of the lowest loop order self-energy diagrams giving rise (from top to bottom) to NP effective masses for quarks, Tera-quarks and $W$. Blobs represent the insertion of the appropriate NP operators among those displayed in equations from~(\ref{OSS}) to~(\ref{OFFA}), and boxes the insertion of the Wilson-like vertices necessary to close the loops. The vertical dotted line means amputation of the external leg.}}
\label{fig:fig6}
\end{figure}

In sect.~\ref{sec:NPMG} we have sketched the calculation of the quark mass. The calculation of the Tera-quark mass exactly parallels what we did there. As for the $W$ mass, from the kind of diagrams displayed in the bottom panel of fig.~\ref{fig:fig6} we find 
\begin{eqnarray}
\hspace{-.8cm}&& M_W^2 \, \propto \,g_w^2 \alpha_{T}^2\Lambda_{T} \, (b^2)^4 \!\int^{\pi/b}\!\frac{d^4 k}{(2\pi)^4} \int^{\pi/b}\!\frac{d^4 \ell}{(2\pi)^4} \int^{\pi/b}\!\frac{d^4 q}{(2\pi)^4}\, \frac{1}{k^2\!+\!m^2_{\zeta_0}} \label{MW1}\\ 
\hspace{-.8cm}&& {\rm tr}\Big[\gamma \!\cdot\! (q\!-\!k) \frac{1}{\gamma\! \cdot \!(q\!-\!k)} q_\mu\frac{1}{\gamma\! \cdot \! q} \gamma \! \cdot \! (q\!-\!\ell) \frac{1}{\gamma \! \cdot \! (q\!-\!\ell)}(q\!-\!\ell)_\mu \frac{1}{\gamma\!\cdot\!q}\Big] \frac{1}{\ell^2\!+\!m^2_{\zeta_0}}\sim 
g_w^2 \alpha_{T}^2 \Lambda_T^2 \, . \nn
\end{eqnarray}
After a few obvious simplifications, one sees that a finite result emerges, like in the case of $m_q$, because the $b^{-8}$ three-loop UV-power divergency is exactly compensated by the factor $b^{8}$ coming from the insertions of the Wilson-like vertices and the NP operators~(\ref{OFFG}), at the end yielding an UV-finite square $W$ mass of O($g_w^2 \alpha_T^2\Lambda_T^2$). 

From the observation that the NP building blocks appearing in the $W$ mass diagrams in the fourth panel of fig.~\ref{fig:fig6} are among those occurring in the second and third panel, we conclude that the dynamical $W$ boson mass is generated by the same NP mechanism that gives mass to fermions and Tera-fermions. 

As we remarked at the end of sect.~\ref{sec:NPMG}, to be complete we should have drawn in each diagram of fig.~\ref{fig:fig6} also the field $U$, exiting from each Wilson-like term. We did not do so because the presence of $U$ is not relevant for what concerns the calculation of the NP effective masses (it is enough to set $U$ equal to the identity). Reinserting the whole dependence upon $U$ is, however, necessary to give rise to the fully $\chi_L\times\chi_R$ invariant mass operators of fermions and Tera-fermions as well as the gauge invariant structure related to the $W$ mass of eq.~(\ref{GWQCNG}) below.

\subsubsection{The $\zeta_0$ propagator}
\label{sec:ZZP}
 
We need to conclude the analysis of these NP mass estimates answering a question that can naturally arise concerning the precise expression of the $\zeta_0$ propagator in the diagrams of fig.~\ref{fig:fig6}. In fact, one might wonder what do we really mean by ``$\zeta_0$ propagator'', given the fact that the critical value of $k_b$ in eq.~(\ref{LKINWQ}) was fixed to precisely cancel the scalar kinetic term against the similar operator with which the Wilson-like terms mix. The cancellation is pictorially represented in fig.~\ref{fig:fig4c} at the lowest loop order. We clarify this important issue in Appendix~\ref{sec:APPC} by disentangling the delicate interplay between the critical and the $b\to 0$ limit in determining the effective expression of the critical $\zeta_0$ propagator. 

\subsection{The NP masses of elementary particles}
\label{sec:NPM}

From the analysis of the building blocks entering the diagrams in fig.~\ref{fig:fig6} one can determine the lowest loop order parametric expression of the effective NP quark, Tera-quark and $W$ mass, finding
\beqn
\hspace{-.6cm}&&m_q=C_q\, \Lambda_{T}\, ,\qquad \qquad \,\,\,C_q={\mbox{O}}(g_s^2\,\alpha_s) \label{MQ1}\\
\hspace{-.6cm}&&m_Q=C_Q \,\Lambda_{T}\, ,\qquad \qquad \,\,C_Q={\mbox{O}}(g_T^2\,\alpha_T,\alpha_T^2,g_s^2\,\alpha_s,g_s^2\,\alpha_T) \label{MQ2}\\
\hspace{-.6cm}&&M_W= C_w\,\Lambda_{T}\, ,\qquad \qquad \,\,C_w=g_w\,c_w\, ,\qquad c_w={\mbox{O}}(\alpha_T) \, , \label{MW}
\eeqn
where the specific choice between $g^2$ and $\alpha$ we made in the coefficients $C_q, C_Q$ comes from taking the product of the factors indicated within the gray blob times either the $\alpha_s$ or the $\alpha_T$ accordng to whether the particle running in the external loop is a SM or a Tera-dof (see Appendix~B in~(II)). 

As for $C_w^2$ (it is more convenient to refer to $M_W^2$), the $\alpha_T^2$ factor coming from the gray blobs, get simply multiplied by a $g_w^2$ factor, because there is no weak interaction loop in the last panel of fig.~\ref{fig:fig6}, in other words, because the weak couplings are external factors and do not pick up the $1/4\pi$ loop-factor.

Considerations of this kind will be important in the sect.~(3.2) of~(II) where, after extending the model to encompass hypercharge interactions introducing leptons and Tera-leptons, we address some interesting phenomenological questions providing a crude estimate of $\Lambda_T$ as well as of some elementary particle mass ratios. Already at the level of the analysis presented above, we can conclude that eqs.~(\ref{MQ1})-(\ref{MW}) allow identifying the EW scale as (a fraction of) the physical parameter $\Lambda_T$, associated to a new interaction. 
 
\subsection{The critical QEL in the NG phase}
\label{sec:CNGP}

We now discuss the form of the QEL which is expected to describe the physics of the renormalized critical model~(\ref{SULLWQ}) in the NG phase at momenta well below the UV cutoff scale but above $\Lambda_T$, i.e.\ in the range $b^{-2}\gg p^2 > \Lambda_T^2$. Following the same line of arguments we developed in sect.~\ref{sec:QNGC} to derive eq.~(\ref{GW}), we get for the $d\leq 4$ piece of the QEL the expression
\beqn
\hspace{-.3cm}&&{\Gamma}_{4\,cr}^{NG}(q,Q;A,G,W;\Phi)=
\frac{1}{4}\Big{(}F^A\!\cdot\! F^A+ F^G\!\cdot\! F^G+F^W\!\cdot \!F^W\Big{)}+\nn\\ 
\hspace{-.3cm}&&+\big{(}\bar q_L \,\,/\!\!\!\! {\cal D}^{WA} q_L+\bar q_R \,\,/\!\!\!\!  {\cal D}^{A}q_R\big{)}+C_q\Lambda_T\,\big{(} \bar q_L U q_R+\bar q_R U^\dagger q_L\big{)}+\nn\\
\hspace{-.3cm}&&+\big{(}\bar Q_L\,\,/\!\!\!\! {\cal D}^{WAG} Q_L\!+\!\bar Q_R\,\,/\!\!\!\! {\cal D}^{AG} Q_R\big{)}+C_Q\Lambda_T\,\big{(} \bar Q_L U Q_R\!+\!\bar Q_R U^\dagger Q_L \big{)}\!+\nn\\
\hspace{-.3cm}&&+\frac{1}{2}c_w^2\Lambda_T^2{\tr}\big{[}({\cal D}\,^{W}_\mu U)^\dagger {\cal D}^{W}_\mu U\big{]} \, ,\label{GWQCNG}
\eeqn 
where (compare with the expression~(\ref{PHPH2}))
\beq
U=\exp[i\vec\tau\,\vec\zeta/c_w\Lambda_T] \, .\label{NGF}
\eeq
In eq.~(\ref{GWQCNG}) we have included $d\leq 4$ operators with up to two derivatives. Actually there exists a further $d=4$ operator invariant under $\chi_L\times\chi_R$, namely ${\tr}\big{[} ({\cal D}^W_\mu U)^\dagger {\cal D}^W_\mu U ({\cal D}^W_\nu U)^\dagger {\cal D}^W_\nu U \big{]}$ which has, however, four derivatives. This term will not impact in an essential way on our considerations in sect.~\ref{sec:ITDF} where we compare the SM Lagrangian with the expression of the LEEL resulting upon integrating out the (heavy) Tera-dof's. It will contribute higher order weak corrections to the LEEL coefficients and/or higher dimensional $d > 4$ operators scaled by powers of $\Lambda_T$. 

In Appendix~\ref{sec:APPB} we justify the form of the expression~(\ref{GWQCNG}) showing that, despite the fact that also the term $\Lambda_T R \, \tr[({\cal D}_\mu^W U)^\dagger {\cal D}^W_\mu U]$ ($R$ is the singlet radial scalar defined in eq.~(\ref{PHPH})), is invariant under $\chi_L\times\chi_R$, it cannot appear in ${\Gamma}_{4\,cr}^{NG}$. The proof of this statement is based on the decoupling theorem~\cite{Appelquist:1974tg} which holds in the critical limit, based to the fact that in the QEL the effective field $\zeta_0$ can be shown to get an infinite mass and decouple. 

Naturally at $d>4$ there will obviously be infinitely many other operators contributing to ${\Gamma}_{cr}^{NG}$, scaled by larger and larger inverse powers of $\Lambda_T$. A few examples are 
{\small \beqn
\hspace{-.8cm}&& \frac{1}{\Lambda_T} \bar Q_L {\overleftarrow {\cal D}}\,_\mu^{AGW} \! U {\cal D}_\mu^{AG} Q_R\!+\!{\mbox{hc}}\, , \,\,
\frac{1}{\Lambda_T}  \bar q_L {\overleftarrow {\cal D}}\,_\mu^{AW} \!U {\cal D}_\mu^{A} q_R\!+\!{\mbox{hc}}\, , \,\,
\frac{1}{\Lambda_T^2}(\bar Q_L UQ_R) (\bar Q_L U Q_R)\!+\!{\mbox{hc}} \nn \\
\hspace{-.8cm}&& \frac{1}{\Lambda_T^2}(\bar q_L U q_R) (\bar q_L U q_R) \!+\!{\mbox{hc}}\, , \quad
\frac{1}{\Lambda_T^4} (\bar Q_L  {\overleftarrow {\cal D}}\,_\mu^{AGW}  U Q_R) (\bar q_L {\overleftarrow {\cal D}}\,_\mu^{AW} \! U q_R) \!+\!{\mbox{hc}} \, ,
\quad \ldots \, .
\label{DQ}
\eeqn}

\subsection{Transversality of the $W$ polarization amplitude}
\label{sec:TWP}

As in the SM the Goldstone bosons $\zeta_i,i=1,2,3$ (see eq.~(\ref{NGF})) are eaten up to give the longitudinal dof's of the massive weak bosons. In Appendix~\ref{sec:APPD} we illustrate how the transversality property of the $W$ polarization amplitude (which in the end is a direct consequence of the SU(2)$_L$ gauge symmetry) is realized in this model. The argument is not totally trivial because in the present situation the $W$ mass is not generated at tree level by the standard Higgs mechanism, but rather arises from NP spontaneous chiral symmetry breaking effects and virtual particle exchanges like we see in the lowest panel of fig.~\ref{fig:fig6}.

\section{Integrating out Tera-degrees of freedom and the SM}
\label{sec:ITDF}

We show in this section that, upon integrating out the (heavy) Tera-dof's, in the NG phase the resulting LEEL of the critical model~(\ref{SULLWQ}) closely resembles the SM Lagrangian.\ The argument rests on the key conjecture that the 125~GeV resonance detected at LHC~\cite{Aad:2012tfa,Chatrchyan:2012xdj} is a $W^+W^-/ZZ$ composite state, bound by Tera-particle exchanges, and not an elementary particle. A non-relativistic argument as well as results based on a simplified Bethe--Salpeter approach, supporting this hypothesis, are presented in~(II). 

This state, which we shall denote by $h$, is (assumed to be) a singlet under all the symmetries of the theory. Since its mass is (experimentally found to be) $\ll\Lambda_T$, it must be included in the LEEL of the theory, valid for (momenta)$^2\ll\Lambda_T^2$. In these kinematical conditions the most general LEEL invariant under the symmetries of the critical model~(\ref{SULLWQ}), in particular under the $\chi_L \times \chi_R$ transformations, and including $h$, takes the form~\cite{Buchalla:2013rka,Brivio:2015hua}
\beqn
\hspace{-.8cm}&&{\cal L}_{LE}^{NG}(q;A,W;U,h)=
\frac{1}{4}\Big{(}F^A\!\cdot\! F^A+F^W\!\cdot\! F^W\Big{)}
+\big{(}\bar q_L \,\,/\!\!\!\! {\cal D}^{AW} q_L+ \bar q_R \,\,/\!\!\!\!  {\cal D}^{A} q_R\big{)} 
+ \nn \\
\hspace{-.8cm}&&\quad +\, (y_q h + k_q k_v) \,\big{(} \bar q_L U q_R+\bar q_R U^\dagger q_L \big{)}+\label{VLELA}\\
\hspace{-.8cm}&&\quad +\frac{1}{2}\partial_\mu h\partial_\mu h +\frac{1}{2}(k_v^2+2k_v k_1 h+k_2h^2){\tr}\big{[} ( {\cal D}\,^{W}_\mu U)^\dagger {\cal D}^{W}_\mu U\big{]} +\widetilde {\cal V}(h)+ \ldots \, , \nn
\eeqn
where dots represent $\tilde\chi_L\times\tilde\chi_R$ violating operators of dimension $d>4$, like for instance 
\beqn
 &&\bar q_L {\overleftarrow {\cal D}}\,_\mu^{AW} U {\cal D}_\mu^{A} q_R+{\mbox{hc}}\, , \qquad d=5 \label{OP5}\\
&& (\bar q_L U q_R) \,(\bar q_R U^\dagger q_L)\, , \qquad\quad d=6 \, .\label{OP6}
\eeqn
They appears in the LEEL multiplied by inverse powers of the fundamental scale $\Lambda_T$ ($\Lambda_T^{-1}$ and $\Lambda_T^{-2}$, respectively). We note that the $1/\Lambda_T$ expansion entailed by the integration over Tera-dofs provides a physical interpretation and a theoretical basis for the occurrence of the higher dimensional operators which are standardly introduced to describe possible new physics beyond the SM.

The scalar potential $\widetilde{\cal V}(h)$ comprises the cubic and quartic self-interactions of the $h$ field, as well as the $h$ mass term, $m_h^2h^2/2$. The $k_v, k_1, k_2, y_q, k_q$ coefficients and the $\widetilde{\cal V}$-couplings are parameters that need to be fixed by matching onto the underlying (renormalizable and unitary) fundamental critical theory~(\ref{SULLWQ}).

Tree-level matching of ${\cal L}_{4,LE}^{NG}$ in eq.~(\ref{VLELA}) with the QEL $\Gamma^{NG}_{4\, cr}$ in eq.~(\ref{GWQCNG}) requires for the masses of quarks and $W$ (see eqs.~(\ref{MQ1}) and~(\ref{MW})) the identifications
\beqn 
&&m_q= C_q \Lambda_T= k_q k_v \, ,\qquad M_W = g_w\, c_w \Lambda_T = g_w k_v \, ,\label{M2}
\eeqn
while the unitarity of the mother theory~(\ref{SULLWQ}) implies the constraints~\cite{Lee:1977eg,Romao:2016ien}
\beq 
y_q=k_q  \, , \qquad k_1 = k_2 = 1 \, .\label{UNIR}
\eeq
The above relations are expected to hold up to small loop effects controlled by the couplings $g_w$ and $y_q$. Neglecting these corrections, one recognizes that, with the exception of the scalar potential $\widetilde{\cal V}(h)$, precisely the combination
\beq
\Phi \equiv (k_v+h) U \label{PHIDEF}
\eeq
enters the $d\leq 4$ part of ${\cal L}_{LE}^{NG}$ (see eq.~(\ref{VLELA})). The latter can thus be rewritten in the suggestive form
\beqn
\hspace{-.8cm}&&{\cal L}_{4,LE}^{NG}(q;A,W;U,h) = 
\frac{1}{4}\Big{(}F^A\!\cdot\! F^A+F^W\!\cdot\! F^W \Big{)} 
+\big{(}\bar q_L \,\,/\!\!\!\! {\cal D}^{AW} q_L+ \bar q_R \,\,/\!\!\!\!  {\cal D}^{A} q_R\big{)} + 
 \nn \\
\hspace{-.8cm}&&\quad +y_q\,\big{(} \bar q_L \Phi q_R+\bar q_R \Phi^\dagger q_L \big{)}+\frac{1}{2}{\tr}\big{[} ( {\cal D}\,^{W}_\mu \Phi)^\dagger {\cal D}^{W}_\mu \Phi\big{]} + \widetilde{\cal V}(h)\, . \label{VLELAMS}
\eeqn
From eq.~(\ref{VLELAMS}) we see that (up to corrections suppressed as O($\alpha_w$) or O($y_q^2/4\pi)$ which we have not included) ${\cal L}_{4,LE}^{NG}$ looks very much like the SM Lagrangian (more precisely, like the Lagrangian of the oversimplified version of the SM where the existence of families, hypercharge interactions, leptons and weak isospin splitting is ignored). In particular, we see that just like it happens in the case of the Higgs mechanism in the SM, the effective Yukawa coupling of $h$ to fermions is given by (see eqs.~(\ref{M2}) and~(\ref{UNIR}))
\beq
y_q=k_q=\frac{m_q}{k_v} =\frac{m_q}{c_w\Lambda_T} \, ,
\label{YKSM}
\eeq
where $k_v=c_w\Lambda_T$ is what in the SM is the Higgs field vev. This simple analysis shows that like in the SM, also here the Yukawa coupling is proportional to the fermion mass. 

There are, however, two important differences between the LEEL of our model and the SM Lagrangian that we must point out. First of all, the proportionality factor between the Yukawa coupling~(\ref{YKSM}) and the fermion mass is not in our hands, rather it is completely fixed by the NP dynamics. Secondly, since the scalar potential $\widetilde{\cal V}(h)$ is supposed to describe, besides the mass, the self-interactions of the (composite) $h$ field, there is no reason why it should have the same form as the SM Higgs potential. This implies that, even if it yields $m_h \!\simeq \!125$~GeV, differences with respect to the case of the SM may well appear in the trilinear and quadrilinear $h$ self-couplings. 

\section{Universality}
\label{sec:UNIV}

A key point that needs to be thoroughly discussed, if we want to put the present approach on a firm conceptual basis and make it useful for phenomenology, is to what extent the NP mass formulae we have derived above are ``universal'', or in other words to what extent their expression depends on the specific form of the ``irrelevant'' $d > 4$ Wilson-like chiral breaking terms that one decides to introduce in the fundamental Lagrangian. 

It turns out that the leading (lowest) power of the gauge coupling dependence of the $C$ coefficients in eqs.~(\ref{MQ1}), (\ref{MQ2}) and~(\ref{MW}) depends on the operator dimension of the Wilson-like terms entering the diagrams of fig.~\ref{fig:fig6}. One can prove that generically the higher is the dimension of the Wilson-like operator the larger will be the leading (lowest) power of the gauge coupling controlling the behaviour of the $C$ coefficients.

Let us then start by analyzing the case in which associated to each fermion species there is a single Wilson-like operator of dimension $d=6$. We prove in Appendix~C of~(II) that in the case of a theory with only one family of quarks and weak interactions (but neither Tera-quarks nor leptons), physics is $\rho$ independent. If other fermions are present, be them quarks belonging to other families, Tera-particles or leptons, observables will only depend on the ratios of the various $\rho$ parameters. 

Notice that in the case numerically studied in ref.~\cite{Capitani:2019syo}, where weak interactions are absent, the quark mass is instead proportional to $\rho^2$. The reason is that the scalar kinetic term is invariant under $\tilde\chi_L\times\tilde\chi_R$ transformations. Hence there is no condition constraining the factor in front of it which is then taken to be equal to unit. On the contrary, in the presence of weak interactions one also needs to introduce and tune an extra coefficient, $k_b$, in front of the scalar kinetic term, because the latter is not invariant under $\tilde\chi_L\times\tilde\chi_R$ transformations. The $k_{b}$ critical value turns out to be proportional to $\rho^2$, as one can see from eq.~(\ref{KCR1}). This yields a compensation of $\rho$ factors in the self-energy diagrams of fig.~\ref{fig:fig6}}, making the quark and the $W$ mass $\rho$ independent if there is only one fermion species or dependent on $\rho$ ratios if we have more than one fermion species.

If the chiral breaking Wilson-like term associated to a fermion is a linear combination of operators of different dimensions, the situation is a bit more complicated to analyze because of mixing. However, one can say the following. It is always the dimension of the Wilson-like operator of lowest dimension that determines the leading (smallest) value of the power of the gauge coupling in the coefficients of the NP mass formulae~(\ref{MQ1}), (\ref{MQ2}) and~(\ref{MW}). Higher and higher dimensional Wilson-like operators affect terms of higher and higher order in the gauge coupling power expansion. 

\subsection{Rescuing universality?}
\label{sec:RU}

At first sight the dependence of the value of the NP masses on the precise form of the chiral breaking Wilson-like terms that we have outlined above might appear as a blunt violation of universality. As we have said above, in fact, the $C$ coefficients in the eqs.~(\ref{MQ1}), (\ref{MQ2}) and~(\ref{MW}) depend not only on the value of the (ratios of the) $\rho$ parameters, but they also have a gauge coupling power dependence correlated with the operator dimension of the various Wilson-like terms in play. 

Let us examine the two issues, $\rho$ dependence and operator dimension,  separately. As for the $\rho$ dependence, the problem could be mitigated or even eliminated by conjecturing that some symmetry exists which, putting constraints on the $\rho$'s, restricts the variability range of their ratios. For instance, in the case where all Wilson-like terms are $d=6$ operators, an extreme but appealing situation would the one in which all the $\rho$'s are equal because of some GUT symmetry. In this case the whole $\rho$ dependence would completely drop out from physical observables. 

Concerning the dependence on the dimensions of the Wilson-like operators, which, as we said, directly impacts on the parametric gauge coupling power dependence of the NP-ly generated masses, the situation may not be as bad as it looks. On the contrary, in our opinion the liberty in the choice of the dimension of the chiral breaking Wilson-like terms might give us an unexpected handle to understand flavour. The idea is to interpret the family mass hierarchy (from heavy to light) as related to Wilson-like terms of increasing dimensions. In this way we may hope to get flavour as an emerging quantum number that identifies different families with a mechanism somehow related to the ``geometry'' of the UV completion of the theory. Possibly also weak isospin breaking could be understood along these lines. Interesting examples where ideas of this kind are exploited for phenomenology are presented in sect.~5 of~(II).

\section{Conclusions and Outlook}
\label{sec:CAO}

In this paper we have shown that, as a viable alternative to the Higgs scenario, a recently discovered~\cite{Frezzotti:2014wja,Capitani:2019syo} NP field theoretical mechanism capable of generating masses for elementary fermions can also provide a mass to the $W$ when the model is extended to include weak interactions. Such a NP phenomenon takes place in strongly interacting theories where the fermion chiral symmetry (or an appropriate extension of it in the presence of weak interactions), broken at the UV cutoff level, is recovered at low energy owing to the tuning of certain Lagrangian parameters. 

As a matter of fact the existence of this NP mass generation mechanism was already noticed in WLQCD, like the compilation of $m_{cr}$ data as function of the lattice spacing reported in fig.~1 of~\cite{Frezzotti:2014wja} shows. However, no use of this NP feature was ever made for the purpose of giving mass to quarks in QCD, because the breaking of chirality induced by the lattice regularization (due to the presence of the Wilson term) has the well-known side effect of generating a linearly divergent term in the mass of the fermion that completely hide any finite underlying NP mass contribution. The standard lattice procedure is thus to subtract the whole $m_{cr}\bar q q$ operator, while at the same time adding by hand an explicit mass term for the quarks.\ The peculiar property of the kind of models discussed in the present paper is that the exact $\chi_L\times\chi_R$ symmetry precisely forbids the existence of linearly divergent quantum corrections to fermion masses, thus opening the possibility of exploiting possible NP finite effects related to the spontaneous breaking of chiral symmetry to provide mass terms to elementary particles.

The approach we are advocating in the present work, which extends ideas already proposed in ref.~\cite{Frezzotti:2014wja}, has also the virtue of offering interesting hints towards the solution of some of the conceptual problems left open by the current formulation of the SM.
\begin{enumerate}
\item First of all, owing to the fact that in our model there is no elementary Higgs, there isn't anymore a ``Higgs mass fine tuning problem'', hence no need to explain why the Higgs mass is so much smaller than the Planck scale~\cite{Susskind:1978ms,THOOFT}. Naturally it remains to explain why the weak force is so much stronger than the gravitational one. In this framework, lacking a fundamental scalar, the whole issue of the stability of the EW vacuum~\cite{Cabibbo:1979ay,Ellis:2009tp,Alekhin:2012py,Salvio:2015cja} may need to be revised. 

\item Secondly, unlike the SM, masses are not free parameters, rather they are NP-ly determined by the dynamics of the theory~\footnote{For completeness, we remark that, as we shall show in~(II) (see eqs.~(2.7) and~(2.8)), in the present approach neutrinos are exactly massless because with the standard hypercharge assignments we are assuming for SM fermions (see Table~1 in ref.~\cite{Frezzotti:2016bes} or in Appendix~A of~(II)) the right-handed neutrino Weyl component is completely sterile.}. 

\item Finally, since the exact $\chi_L\times\chi_R$ symmetry protects  elementary particle masses from quantum power divergencies, the former are ``naturally small'', i.e.\ ${\mbox{O}}(\Lambda_{\rm RGI})$, and indeed  the massless critical theory enjoys an enhanced symmetry of chiral nature. From this point of view the present approach to mass generation complies with the 't Hooft criterion of naturalness~\cite{THOOFT}.
\end{enumerate}

We have also seen that to cope with the magnitude of the top and $W$ mass, a super-strongly interacting sector, gauge invariantly coupled to SM matter, must be conjectured to exist, in order for the full theory, including SM matter and Tera-particles, to have an RGI scale $\Lambda_{\rm RGI}\!\sim\!\Lambda_T\!\gg\!\Lambda_{QCD}$ of the order of a few TeVs. We have proved in this paper that the simplest model, introduced in~\cite{Frezzotti:2014wja} and endowed with the above NP mass generation mechanism, can indeed be extended to incorporate weak interactions and the conjectured Tera-dof's. In a similar way one could include leptons and the hypercharge interaction. This further step is discussed in~(II).

As for the relation of the present scheme to the SM, we have observed in sect.~\ref{sec:ITDF} that upon integrating out the (heavy) Tera-dof's, under the assumption that a ``light'' (compared to the TeV scale, $\Lambda_T$) $W^+W^-/ZZ$ composite scalar singlet state bound by Tera-particle exchanges gets formed (that we conjecture should be identified with the 125~GeV resonance discovered at LHC~\cite{Aad:2012tfa,Chatrchyan:2012xdj}), one ends up with a LEEL valid for (momenta)$^2\ll \Lambda_T^2$, which, ignoring small perturbative corrections, resembles very much the SM Lagrangian, except possibly for the effective trilinear and quadrilinear self-couplings of the composite scalar.

\renewcommand{\thesection}{A} 
\section{Symmetries, currents and tuning}  
\label{sec:APPA}

In this Appendix we provide a derivation of the criticality conditions which allow enforcing the invariance of the Lagrangian~(\ref{SULLWQ}) under the $\tilde\chi_L\times\tilde\chi_R$ transformations~(\ref{GTWTQ})-(\ref{GTCTQ}).

The exactly conserved currents associated with the transformations $\chi_L\times\chi_R$ of eq.~(\ref{CHILN}) have the expression ($i=1,2,3$)
\beqn
\hspace{-.6cm}&&J_\mu^{L\,i}= K_\mu^i+\bar q_L\gamma_\mu\frac{\tau^i}{2}q_L+\bar Q_L\gamma_\mu \frac{\tau^i}{2} Q_L-\frac{k_b}{4}{\tr}\big{[}\Phi^\dagger\frac{\tau^i}{2}{\cal D}^W_\mu\Phi-(\Phi\overleftarrow{\cal D}\,^W_\mu)^\dagger\frac{\tau^i}{2}\Phi\big{]}+\nn\\
\hspace{-.6cm}&&\quad-\frac{b^2}{2}\rho_q\,\big{(} \bar q_L \frac{\tau^i}{2}\Phi {\cal D}^A_\mu q_R-\bar q_R \overleftarrow{\cal D}\,^A_\mu \Phi^\dagger \frac{\tau^i}{2} q_L\big{)}+\nn\\
\hspace{-.6cm}&&\quad-\frac{b^2}{2}\rho_Q\,\big{(} \bar Q_L \frac{\tau^i}{2}\Phi {\cal D}^{AG}_\mu Q_R-\bar Q_R \overleftarrow{\cal D}\,^{AG}_\mu \Phi^\dagger \frac{\tau^i}{2} Q_L\big{)}\, , \label{CCLQ}\\
\hspace{-.6cm}&&J_\mu^{R\,i}\!=\! \bar q_R\gamma_\mu\frac{\tau^i}{2}q_R+\bar Q_R\gamma_\mu \frac{\tau^i}{2} Q_R-\frac{k_b}{4}{\tr}\big{[}(\Phi\overleftarrow{\cal D}\,^W_\mu)^\dagger \Phi\frac{\tau^i}{2}-\frac{\tau^i}{2}\Phi^\dagger ({\cal D}^W_\mu\Phi)\big{]}+\nn\\
\hspace{-.6cm}&&\quad-\frac{b^2}{2}\rho_q\,\big{(}\bar q_R \frac{\tau^i}{2}\Phi^\dagger {\cal D}\,^{AW}_\mu q_L\!-\!\bar q_L \overleftarrow{\cal D}\,^{AW}_\mu \Phi \frac{\tau^i}{2} q_R\big{)}+\nn\\
\hspace{-.6cm}&&\quad-\frac{b^2}{2}\rho_Q\,\big{(}\bar Q_R \frac{\tau^i}{2}\Phi^\dagger {\cal D}\,^{AGW}_\mu Q_L\!-\!\bar Q_L \overleftarrow{\cal D}\,^{AGW}_\mu \Phi \frac{\tau^i}{2} Q_R\big{)}\, ,
\label{CCRQ}
\eeqn
where
\beq
K_\mu^i=g_w\tr\Big{(}[W_\nu,F^W_{\mu\nu}]\frac{\tau^i}{2}\Big{)}\, . \label{KWI}
\eeq
We stress that the conserved currents, $J^{L\,i}$, to which weak bosons are coupled, do not coincide with the conserved left-handed currents of the SM because of the presence of Tera-particle contributions and of the terms originating from the variation of the $d=6$ Wilson-like operators.

The key observation here is that, unlike the case where weak interactions are absent, in the Lagrangian~(\ref{SULLWQ}), neither the Yukawa and Wilson-like operators, nor the $\Phi$ kinetic term are invariant under the $\tilde \chi_L\times\tilde \chi_R$ transformations as the latter do not act on the scalar field (see eqs.~(\ref{GTWTQ})-(\ref{GTCTQ})). The (non-conserved) currents associated to the (chiral) $\tilde \chi_L\times\tilde \chi_R$ transformations read
\beqn
\hspace{-.6cm}&&\tilde J_\mu^{L\,i}= K_\mu^i+\bar q_L\gamma_\mu\frac{\tau^i}{2}q_L+\bar Q_L\gamma_\mu \frac{\tau^i}{2} Q_L+\nn\\
\hspace{-.6cm}&&\quad-\frac{b^2}{2}\rho_q \big{(} \bar q_L \frac{\tau^i}{2}\Phi {\cal D}^A_\mu q_R-\bar q_R \overleftarrow{\cal D}\,^A_\mu \Phi^\dagger \frac{\tau^i}{2} q_L\big{)}+\nn\\
\hspace{-.6cm}&&\quad-\frac{b^2}{2}\rho_Q\,\big{(} \bar Q_L \frac{\tau^i}{2}\Phi {\cal D}^{AG}_\mu Q_R-\bar Q_R \overleftarrow{\cal D}\,^{AG}_\mu \Phi^\dagger \frac{\tau^i}{2} Q_L\big{)}\, , \label{CCLQQ}\\
\hspace{-.6cm}&&\tilde J_\mu^{R\,i}= \bar q_R\gamma_\mu\frac{\tau^i}{2}q_R+\bar Q_R\gamma_\mu \frac{\tau^i}{2} Q_R+\nn\\
\hspace{-.6cm}&&\quad-\frac{b^2}{2}\rho_q \big{(}\bar q_R \frac{\tau^i}{2}\Phi^\dagger {\cal D}^{AW}_\mu q_L\!-\!\bar q_L \overleftarrow{\cal D}\,^{AW}_\mu \Phi \frac{\tau^i}{2} q_R\big{)}+\nn\\
\hspace{-.6cm}&&\quad-\frac{b^2}{2}\rho_Q\,\big{(}\bar Q_R \frac{\tau^i}{2}\Phi^\dagger {\cal D}^{AGW}_\mu Q_L\!-\!\bar Q_L \overleftarrow{\cal D}\,^{AGW}_\mu \Phi \frac{\tau^i}{2} Q_R\big{)}\, .
\label{CCRQQ}
\eeqn
They differ from the conserved currents, $J^{L\, i}_\mu$ and $J^{R\, i}_\mu$, only because the terms bilinear in the scalar field coming from variation of the $\Phi$-kinetic term (proportional to $k_b$) are absent in~(\ref{CCLQQ}) and~(\ref{CCRQQ}).

\subsection{Enforcing invariance under the chiral $\tilde\chi_L\times\tilde\chi_R$ transformations}
\label{sec:ENFCHTILDE}

Generalizing the strategy proposed in~\cite{Frezzotti:2014wja}, restoration of chirality (up to O($b^2$) cutoff effects) is obtained by appropriately tuning the Yukawa couplings $\eta_q$ and $\eta_Q$ and the coefficient $k_b$ to critical values determined by enforcing the conservation of the currents associated with the $\tilde\chi_L\times\tilde\chi_R$ transformations. 
 
In order to identify under which conditions the (chiral) $\tilde\chi_L\times\tilde\chi_R$ transformations~(\ref{GTWTQ}) and~(\ref{GTCTQ}) can become an invariance of the Lagrangian~(\ref{SULLWQ}) we start by writing down the Schwinger--Dyson equations (SDEs) stemming from the $\tilde\chi_L\times\tilde\chi_R$ transformations~\footnote{We refrain from calling Ward--Takahashi identities (WTIs) eqs.~(\ref{CTLTIRCRQ}) and~(\ref{CTRTIRCRQ}) below, because they refer to transformations that are not symmetries of the Lagrangian. For generic values of $\eta_q$, $\eta_Q$ and $k_b$ we prefer to talk of Schwinger--Dyson equations. Only at the critIcal point where the $\tilde\chi_L\times\tilde\chi_R$ currents are conserved one is entitled to talk of WTIs.}. Following the steps outlined in~\cite{Frezzotti:2014wja}, based on the strategy devised in~\cite{Bochicchio:1985xa,Testa:1998ez}, we arrive at the (renormalized) SDEs 
{\beqn													\hspace{-.6cm}&&\partial_\mu \langle Z_{\tilde J}\widetilde J^{L\, i}_\mu(x) \,\hat O(0)\rangle = \langle \widetilde\Delta_{L}^i\hat O(0)\rangle\delta(x) +\label{CTLTIRCRQ}\\
\hspace{-.6cm}&&\quad-(\eta_q-\bar \eta^L_q) \,\langle \Big{(} \bar q_L\frac{\tau^i}{2}\Phi q_R-\bar q_R\Phi^\dagger\frac{\tau^i}{2}q_L \Big{)}(x)\,\hat O(0)\rangle +\nn\\
\hspace{-.6cm}&&\quad-(\eta_Q-\bar \eta^L_Q) \,\langle \Big{(} \bar Q_L\frac{\tau^i}{2}\Phi Q_R-\bar Q_R\Phi^\dagger\frac{\tau^i}{2}Q_L \Big{)}(x)\,\hat O(0)\rangle +\nn\\
\hspace{-.6cm}&&\quad+\frac{i}{2}g_w(k_b \!-\! \bar k_b^L)\langle\tr\!\Big{(}\Phi^\dagger [\frac{\tau^i}{2},\!W_\mu] {\cal D}_\mu^W \Phi \!+\! \Phi^\dagger \overleftarrow {\cal D}\,^W_\mu  [W_\mu,\!\frac{\tau^i}{2}] \Phi\Big{)} (x) \hat O(0)\rangle \!+\! {\mbox O}(b^2)\!+\!\ldots\, ,\nn\\
\hspace{-.8cm}&&\nn\\
\hspace{-.8cm}&&\partial_\mu \langle Z_{\widetilde J}\tilde J^{R\, i}_\mu(x) \,\hat O(0)\rangle = \langle \widetilde\Delta_{R}^i \hat O(0)\rangle\delta(x) +\label{CTRTIRCRQ}\\
\hspace{-.8cm}&&\quad-(\eta_q-\bar\eta^R_q) \,\langle \Big{(} \bar q_R\frac{\tau^i}{2}\Phi^\dagger q_L-\bar q_L\Phi\frac{\tau^i}{2} q_R \Big{)}(x)\,\hat O(0)\rangle +\nn\\
\hspace{-.8cm}&&\quad-(\eta_Q-\bar\eta^R_Q) \,\langle \Big{(} \bar Q_R\frac{\tau^i}{2}\Phi^\dagger Q_L-\bar Q_L\Phi\frac{\tau^i}{2} Q_R \Big{)}(x)\,\hat O(0)\rangle 
+{\mbox O}(b^2)\!+\!\ldots\, ,\nn
\eeqn
where $\hat O(0)$ is a generic local operator and the quantities $\bar\eta^{L,R}_q$, $\bar\eta^{L,R}_Q$ and $\bar k_b^L$ are the mixing coefficients of the $\tilde\chi_{L,R}$ variations of the Wilson-like and Tera-Wilson-like terms with the  variations of the quark and Tera-quark Yukawa operators and the scalar kinetic term, respectively.

The tuning conditions yielding the conservation of the $\tilde\chi_L\times\tilde\chi_R$ currents that determine the values of the parameters $\eta_q$, $\eta_Q$ and $k_b$ of the critical theory then take the form 
\beqn
&&\eta_q-\bar\eta^L_q(\{g\};\eta_q,\eta_Q,\rho_q,\rho_Q;k_b;\mu_0,\lambda_0)=0 \, ,\label{CREQSQ1}\\
&&\eta_q-\bar\eta^R_q(\{g\};\eta_q,\eta_Q,\rho_q,\rho_Q;k_b;\mu_0,\lambda_0)=0 \, ,\label{CREQSQ2}\\
&&\eta_Q-\bar\eta^L_Q(\{g\};\eta_q,\eta_Q,\rho_q,\rho_Q;k_b;\mu_0,\lambda_0)=0 \, ,\label{CREQSQ3}\\
&&\eta_Q-\bar\eta^R_Q(\{g\};\eta_q,\eta_Q,\rho_q,\rho_Q;k_b;\mu_0,\lambda_0)=0 \, ,\label{CREQSQ4}\\
&&k_b\,-\,\bar k_b^L(\{g\};\eta_q,\eta_Q,\rho_q,\rho_Q;k_b;\mu_0,\lambda_0)=0 \, ,\label{CREQSQ5}
\eeqn
where for short we have set $\{g\}=(g_s,g_T,g_w)$. Notice that, unlike what happens in the absence of weak interactions where parity is unbroken, we have for both quarks and Tera-quarks a ``$Left$'' and a ``$Right$'' equation determining the Yukawa couplings. Thus one may wonder whether in the present situation, where parity is not an exact symmetry of the basic Lagrangian, the sets of $Left$ (eqs.~(\ref{CREQSQ1}) and~(\ref{CREQSQ3}) and $Right$ (eqs.~(\ref{CREQSQ2}) and~(\ref{CREQSQ4})) constraints are compatible with each other. In the next subsection we prove that $Left-Right$ compatibility is a consequence of the exact $\chi_L\times\chi_R$ symmetry of the theory. 

\subsubsection{Compatibility between $Right$ and $Left$ tuning conditions}  

In this section we discuss the compatibility of the constraints  from~(\ref{CREQSQ1}) to~(\ref{CREQSQ5}) in the situation where, because of the presence of weak interactions, parity is not an exact symmetry of the basic Lagrangian. We will show that, owing to the exact $\chi_L\times\chi_R$ invariance (plus standard symmetries and dimensionality arguments) the set of $\eta_{f}^L$, $\eta_{f}^L$ ($f=q,Q$) and $k_b$ parameters that satisfies the subset of conditions following from the requirement of, say, $\tilde\chi_L$ symmetry restoration alone, automatically satisfies also the conditions corresponding to $\tilde\chi_R$ symmetry restoration. 

To reduce the argument to its essentials, for simplicity of notations we imagine working in a model where only quarks and $W$'s are present. In this situation the tuning conditions reduce to the following set of equations
\beqn
&&\eta_q-\bar\eta_q^L(\{g\};\eta_q, \rho;k_b;\mu_0,\lambda_0)=0 \, ,\label{CREQSL1}\\
&&\eta_q-\bar\eta_q^R(\{g\};\eta_q, \rho;k_b;\mu_0,\lambda_0)=0\, ,\label{CREQSR1}\\
&&k_b\,-\,\bar k^L_b (\{g\};\eta_q, \rho;k_b;\mu_0,\lambda_0)=0 \, ,\label{CREQS3} 
\eeqn
where $\{g\}=(g_s,g_w)$. Let us start by provisionally identifying as $\eta_{q\,cr}$ and $k_{b\,cr}$ the values obtained by solving, say,  eqs.~(\ref{CREQSL1}) and~(\ref{CREQS3}) which enforce $\tilde\chi_L$ invariance. We notice that a bit away from the critical limit, owing to dimensional considerations, hermiticity and exact $\chi_L\times\chi_R$ invariance, the $d=4$ QEL of the theory in the Wigner phase must have the form  
\beqn
\hspace{-.8cm}&& {\Gamma}_4^{Wig} = \frac{1}{4} \Big{(}F^A\!\cdot\! F^A+F^W\!\cdot \!F^W\Big{)} + 
\Big{[}\bar q_L\,\Dslash^{A,W} q_L+\bar q_R\,\Dslash^{A} q_R\Big{]} + 
\nonumber \\
\hspace{-.8cm}&&\quad+\eta_{q}^{red}\,\big{(} \bar q_L\Phi\, q_R+\bar q_R \Phi^\dagger q_L\big{)}+\frac{k_{b}^{red}}{2}{\tr}\big{[}({\cal D}\,^W_\mu \Phi)^\dagger {\cal D}^W_\mu\Phi\big{]} \, + {\cal V}(\Phi)  \, ,
\label{L4WIG2}
\eeqn
where $\eta_{q}^{red}$ and $k_{b}^{red}$ are the ``reduced'' couplings
\beqn
&&\eta_{q}^{red} =\eta_q-\bar \eta_q^L(\{g\};\eta_q, \rho;k_b;\mu_0,\lambda_0)\, ,\label{YEFF}\\
&&k_{b}^{red} =k_b-\bar k^L_b(\{g\};\eta_q, \rho;k_b;\mu_0,\lambda_0)\, . \label{KEFF}
\eeqn
Setting $\eta_q$ and $k_b$ equal to the solutions of the eqs.~(\ref{CREQSL1}) and~(\ref{CREQS3}), the restoration of the $\tilde\chi_{L}$ symmetry entails the vanishing of the reduced coefficients
\beqn
&&\eta_{q}^{red}\, \stackrel {\eta_q \to \eta_{q\,cr},\, k_b \to k_{b\,cr}} \longrightarrow \, 0 \, ,\label{YEFFV}\\
&&k_{b}^{red}\, \stackrel {\eta_q \to \eta_{q\,cr},\, k_b \to k_{b\,cr}} \longrightarrow \, 0^+ \, . \label{KEFFV}
\eeqn
Looking at the form of eq.~(\ref{L4WIG2}) at $y_{q}^{red}=0$ and $k_{b}^{red} \to 0^+$, it is apparent that the $\tilde\chi_R$ invariance is also automatically restored and the current $\tilde J_\mu^{R,\, i}$ is consequently conserved because the vanishing of $\eta_{q}^{red}$ and $k_{b}^{red}$ makes all chiral breaking terms disappear from the QEL~(\ref{L4WIG2}). This fact in turn implies that the values of $\eta_{q\,cr}$ and $k_{b\,cr}$ determined from the eqs.~(\ref{CREQSL1}) and~(\ref{CREQS3}) will also fulfil eq.~(\ref{CREQSR1}). The extension of this argument to the case of the model considered in the main text, which also includes Tera-fermions and Tera-strong interactions, is straightforward.

\renewcommand{\thesection}{B} 
\section{The QEL of the critical model in the NG phase}  
\label{sec:APPB}

In this Appendix we want to justify the form~(\ref{GWQCNG}) that the QEL of the renormalizable model~(\ref{SULLWQ}) takes in the NG phase at the critical point. In particular we want to prove that, despite the fact that the operator $\Lambda_T R \, \tr[({\cal D}_\mu^W U)^\dagger {\cal D}^W_\mu U]$ is invariant under $\chi_L\times\chi_R$ transformations, it cannot appear in $\Gamma^{NG}_{cr}$ (see eq.~(\ref{GWQCNG})). To this purpose it is convenient to start by examining the structure of the QEL, $\Gamma^{NG}$, slightly away from the critical point. Using the definition introduced in eq.~(\ref{PHPH}), we can write
\beqn
\hspace{-.6cm}&& \Gamma^{NG} \!= \!\Gamma^{NG}_{4\, cr}\! +\! 
\frac{\mu_\Phi^2}{2} R^2\!+\!\frac{\lambda}{4}
R^4\! +\! \frac{1}{2} k_{b}^{red}\Big{(}\partial_\mu R\, \partial_\mu R\! +\! R^2 \tr[({\cal D}_\mu^W U^\dagger) {\cal D}^W_\mu U] \Big{)} \!+\! \nn \\
\vspace{0.1cm}
\hspace{-.6cm}&& + \widetilde c_{\Phi}\, \Lambda_T 
R \, \tr[({\cal D}_\mu^W U)^\dagger {\cal D}^W_\mu U] 
+ \eta_{q}^{red} R \Big( \bar q_L U q_R+\bar q_R U^\dagger q_L \Big) + \nn \\
\vspace{0.4cm}
\hspace{-.6cm}&& + \eta_{Q}^{red} R \Big( \bar Q_L U Q_R+\bar Q_R U^\dagger Q_L \Big)  + \ldots \, ,
\label{FUNGQEL}
\eeqn
where ${\Gamma}_{4\,cr}^{NG}$ is the $d=4$ QEL of the model in the critical limit, given in eq.~(\ref{GWQCNG}). It includes all the NP $\tilde\chi_L \times \tilde\chi_R$ breaking terms of O($\Lambda_T$) and O($\Lambda_T^2$).\ Dots stand for $d>4$ operators describing further NP $\tilde\chi_L \times \tilde\chi_R$ breaking operators that are suppressed by inverse powers of $\Lambda_T$ (like for instance those in eq.~(\ref{DQ})). The ``reduced'' coefficients $\eta_{q}^{red}$, $\eta_{Q}^{red}$ and $k_{b}^{red}$ are defined by the relations (which generalize eqs.~(\ref{YEFF}) and~(\ref{KEFF})) 
\beqn
&&\eta_{q}^{red} = \eta_q - \bar\eta_q^L(\{g\};\eta_q,\eta_Q,\rho_q,\rho_Q;k_b;\mu_0,\lambda_0)\, , \label{REDCOUP1}\\
&&\eta_{Q}^{red} = \eta_Q - \bar\eta_Q^L (\{g\};\eta_q,\eta_Q,\rho_q,\rho_Q;k_b;\mu_0,\lambda_0)\, , \label{REDCOUP2}\\
&&k_b^{red} = k_b - \bar k^L_b(\{g\};\eta_q,\eta_Q,\rho_q,\rho_Q;k_b;\mu_0,\lambda_0)\, ,  \label{REDCOUP3}
\eeqn
respectively, with the critical limit determined by the conditions
\beq
\eta_{q}^{red} = \eta_{Q}^{red} =0\, , \qquad k_{b}^{red}\to 0^+ \, .
\label{CLIM}
\eeq
The purpose of this Appendix is to show that also the coupling $\widetilde c_{\Phi}$ vanishes in the critical limit, implying that the undesired contribution of order $g_w^2 v \Lambda_T$ to $M_W^2$, that would arise from this term, cannot be present, making $M_W$ independent of the vev of the scalar.

Since the QEL is a functional from which one is supposed to directly extract the full quantum information of the model, it is mandatory to normalize canonically the effective scalar field $R$ appearing in eq.~(\ref{FUNGQEL}). Upon introducing
\beq
\Phi_c=\Phi \sqrt{k_{b}^{red}}\, , \qquad
R_c=R \sqrt{k_{b}^{red}}\, , \label{RCAN}
\eeq
with the subscript ``$c$'' standing for ``canonical'', $\Gamma^{NG}$ takes the form
\beqn
\hspace{-.7cm}&& \Gamma^{NG} \!=\! \Gamma^{NG}_{4\, cr} + 
\frac{\mu_\Phi^2}{2 k_{b}^{red}}R_c^2+\frac{\lambda}{4 (k_{b}^{red})^2}R_c^4 +
\frac{1}{2} \Big{(} (\partial_\mu R_c )^2 + R_c^2 \tr[({\cal D}_\mu^W U)^\dagger {\cal D}^W_\mu U] \Big{)} \!+ \nn \\
\vspace{0.1cm}
\hspace{-.7cm}&& + \frac{\widetilde c_{\Phi}}{\sqrt{k_{b}^{red}}} \, \Lambda_T 
R_c  \tr[({\cal D}_\mu^W U)^\dagger {\cal D}^W_\mu U] 
+ \frac{\eta_{q}^{red}}{\sqrt{k_{b}^{red}}} R_c \Big( \bar q_L U q_R+\bar q_R U^\dagger q_L \Big) + \nn \\
\vspace{0.4cm}
\hspace{-.7cm}&& + \frac{\eta_{Q}^{red}}{\sqrt{k_{b}^{red}}} R_c \Big( \bar Q_L U Q_R+\bar Q_R U^\dagger Q_L \Big) +\ldots \, ,
\label{FUNGQELC}
\eeqn
where, recalling $v^2=|\mu_\Phi^2|/\lambda$, we have
\beq
R_c=v_c+\zeta_{0c} \, ,\qquad \zeta_{0c} =\zeta_0\sqrt{k_{b}^{red}}\, , \qquad v_c^2=\frac{|\mu_\Phi^2|}{\lambda} k_{b}^{red}\, .
\label{VQC}
\eeq
From the definition of $R_c$, the expression~(\ref{FUNGQELC}) of $\Gamma^{NG}$ and eq.~(\ref{VQC}), it is apparent that a peculiar non-linear sigma-model is realized in the scalar sector as $k_{b}^{red}\to 0^+$ because
\beq
 v_c \sim \sqrt{k_{b}^{red}} \to 0^+   \, , \quad
 m^2_{\zeta_{0c}}=\frac{2|\mu_\Phi^2|}{k_{b}^{red}}\to +\infty \, .
\label{MZ0}
\eeq
We see that in the critical limit the squared mass of the effective $\zeta_{0c}$ mode is a (real positive) divergent quantity while the vev of $R_c$ vanishes because its effective quartic coupling, $\lambda / 4 (k_{b}^{red})^2$, diverges faster than $m^2_{\zeta_{0c}}$. As for the canonical reduced Yukawa couplings of quarks and Tera-quarks ${\eta_{q}^{red}}/{\sqrt{k_{b}^{red}}}$ and ${\eta_{Q}^{red}}\!/\!{\sqrt{k_{b}^{red}}}$, they can be safely set to zero before taking the limit $k_{b}^{red}\!\to\!0^+$.

We are now ready to show that in the critical limit~(\ref{CLIM}) the coupling $\widetilde c_{\Phi}/ \sqrt{k_{b}^{red}}$ appearing in the QEL~(\ref{FUNGQELC}) is finite, or equivalently that 
\beq
\widetilde c_{\Phi}\sim \sqrt{k_{b}^{red}} \to 0 \, ,
\label{CTO}
\eeq
which validates the expression of $\Gamma^{NG}_{4\, cr}$ given in~(\ref{GWQCNG}) where the term proportional to $\widetilde c_\Phi$ was omitted. 

The key remark on which the proof is based is that, owing to the decoupling theorem~\cite{Appelquist:1974tg}, the QEL $\Gamma^{NG}$ (see eq.~(\ref{FUNGQELC})) must be such as to yield in the critical limit amplitudes with the virtual exchange of one $\zeta_{0c}$ particle that vanish at least like $1/m^2_{\zeta_{0c}}$. To see the implications of this constraint let us for instance consider the $WW\to WW$ scattering amplitude, which receives a tree-level contribution from the exchange of a $\zeta_{0c}$ particle. Taking the $WW\zeta_{0c}$-vertex from the term $\propto \widetilde c_{\Phi}\Lambda_T$ in the second line of~(\ref{FUNGQELC}), one gets
\beq
A(WW\to WW) \propto g_w^2 \frac{\widetilde c_{\Phi} \Lambda_T}{\sqrt{k_{b}^{red}}} \,\frac{1}{s+m^2_{\zeta_{0c}}} \, \frac{\widetilde c_{\Phi}\Lambda_T}{\sqrt{k_{b}^{red}}}\, g_w^2\, .
\label{WWAMPL}
\eeq
Since for large $m^2_{\zeta_{0c}}$ and at fixed $s$, the decoupling theorem requires
\beq
A(WW\to WW) \propto (g_w^2 \Lambda_T)^2 \frac{(\widetilde c_{\Phi})^2}{k_{b}^{red}}
\frac{1}{s+m^2_{\zeta_{0c}}}\stackrel {k_{b}^{red} \to 0^+}\longrightarrow  {\mbox{O}}\Big{(}{\frac{1}{m^2_{\zeta_{0c}}}} \Big{)}\, ,
\label{VAN}
\eeq
or faster, it follows that indeed eq.~(\ref{CTO}) must hold, otherwise as $k^{red}_b\to 0$ this contribution to $A(WW\to WW)$, instead of vanishing like $1/m^2_{\zeta_{0c}}$, would diverge. As we said, the vanishing of $\widetilde c_{\Phi}$, that was already accounted for in the expression of ${\Gamma}_{4\,cr}^{NG}$ we gave in eq.~(\ref{GW}), tells us that $M_W$ does not depend on the scalar vev.

A physical consequence of paramount importance of the analysis carried out above is that the (canonically normalized) singlet scalar mode, $\zeta_{0c}$, becomes an infinitely massive field with vanishing vev, decoupled from fermions and gauge bosons, with no dynamics on any physical scale. This in turn implies that there isn't any dependence of physical observables on the scalar quartic coupling, $\lambda_0$ and that, as announced, the $d=4$ QEL of the critical theory in the NG phase is given by the functional ${\Gamma}_{4\,cr}^{NG}$ reported in eq.~(\ref{GWQCNG}).

\renewcommand{\thesection}{C} 
\section{The $\zeta_0$ critical propagator}  
\label{sec:APPC} 

An apparently tricky question is what is the expression of the $\zeta_0$ propagator in the critical limit, in view of the fact that the value of $k_b$ in eq.~(\ref{LKINWQ}) is fixed to precisely cancel the scalar kinetic term against the similar operators with which the Wilson-like terms mix. At the lowest loop order this cancellation is pictorially represented in fig.~\ref{fig:fig4c}. 

To answer this question we need to start from the form of the quadratic part of the scalar field Lagrangian before tuning, which reads
\beq
{\cal L}_{\Phi}^{(2)}=\frac{k_b}{2\delta}{\tr}\big{[}({\cal D}\,^{W}_\mu \Phi)^\dagger{\cal D}^W_\mu\Phi\big{]} +\frac{m_{\zeta_0}^2}{2}{\tr}\big{[} \Phi^\dagger\Phi\big{]} \, ,\label{SCQU}
\eeq
where $\delta$ is a parameter that has been momentarily introduced to canonically normalize $\zeta_0$ (recall $\Phi=(v+\zeta_0)U$).

Looking at fig.~\ref{fig:fig4c}, we can write for the amputated $\zeta_0$-propagator 
\beq
\Pi(p^2)\Big{|}^{1-loop}_{amp}=\frac{1}{\delta}(k_b p^2+m_{\zeta_0}^2) - K(p^2) \, ,
\label{AMP1}
\eeq
where for the 1-loop correction we get 
\beq
K(p^2) = k_0 b^{-2}+k_1 p^2+k_2 b^2p^4+ \ldots
\label{KP2}
\eeq
Inserting this expansion in eq.~(\ref{AMP1}), we obtain
\beqn
&&\Pi(p^2)\Big{|}^{1-loop}_{amp}=\frac{1}{\delta}(k_bp^2+m_{\zeta_0}^2) - k_0 b^{-2}-k_1 p^2- k_2 b^2p^4+ \ldots =\nn\\
&&=\Big{(}\frac{k_b}{\delta}-k_1\Big{)}p^2 +\Big{(}\frac{m_{\zeta_0}^2}{\delta}-k_0 b^{-2}\Big{)} - k_2 b^2p^4 + \ldots  \, .\label{AMP2}
\eeqn
These equations tell us that the tuning condition is 
\beq
k_{b \,cr}=\delta k_1\, ,
\label{KPCR}
\eeq
from which we have 
\beqn
&&\Pi(p^2)\Big{|}^{1-loop}_{amp}=\Big{(}\frac{k_b}{\delta}-k_1\Big{)}p^2 +\Big{(}\frac{m_{\zeta_0}^2}{\delta}-k_0 b^{-2}\Big{)} - k_2 b^2p^4+\ldots=\nn\\
&&=  \frac{k_b-k_{b \,cr}}{\delta}p^2 +\Big{(}\frac{m_{\zeta_0}^2}{\delta}-k_0 b^{-2}\Big{)} - k_2 b^2p^4 + \ldots\, .\label{AMP3}
\eeqn
To canonically normalize the scalar kinetic term we need to take 
\beq
\delta =k_b -k_{b\,cr}	\, .
\label{KPCRR}
\eeq
The canonically normalized, amputated $\zeta_0$-propagator then reads
\beq
\Pi(p^2)\Big{|}^{1-loop}_{amp}=p^2 +\Big{(}\frac{m_{\zeta_0}^2}{k_b -k_{b\,cr}}-k_0 b^{-2}\Big{)} - k_2 b^2p^4 + \ldots \, .
\label{NORM}
\eeq
We now note that the self-energy diagrams (like those shown in fig.~\ref{fig:fig6}) have a finite value which is independent of the $\zeta_0$ mass, if the limit $b\to 0$ is taken before setting $k_b -k_{b\,cr}\to 0$.

The result of this analysis is that the ``effective'' mass of $\zeta_0$ diverges 
in the critical limit, without affecting the value of the self-energy diagrams, provided, we insist, the limit $b\to 0$ is taken before letting $k_b -k_{b\,cr}\to 0$, as we should, and we have actually implicitly done in the analysis presented in Appendix~\ref{sec:APPB} which is based on the continuum-like form of the QEL.

\renewcommand{\thesection}{D}
\section{Transversality of $W$ polarization amplitude}
\label{sec:APPD}

In this Appendix we want to illustrate how the expected transversality property of the $W$ polarization amplitude is realized in the critical limit of the model~(\ref{LWILWQ}) and discuss how the Goldstone fields $\zeta_i, i=1,2,3$ (see eq.~(\ref{NGF})) get eaten up to become the longitudinal $W$ modes~\cite{PESK}. In particular, we will show that the sum of the amputated diagrams displayed in the top panel of fig.~\ref{fig:FIG33} has the expected transverse structure, namely
\beqn
\hspace{-1.8cm}&&\langle \widehat W^i_\mu(p) \widehat W^j_\nu(-p)\rangle \Big{|}^{\rm amp}_{p\to 0}\! \!=\! g_w^2 \langle \widehat J^{L\,i}_\mu(p) \widehat J^{L\,j}_\nu(-p)\rangle \Big{|}_{p\to 0}\!\! \!\to\! c^2_w\Lambda_T^2 g_w^2 \Big{[}\delta_{\mu\nu}\!-\!\frac{p_\mu p_\nu}{p^2}\Big{]}\delta_{ij}\, .
\label{TPF}
\eeqn
In eq.~(\ref{TPF}) we have dropped an irrelevant $\delta(0)$ factor and indicated by $\widehat W^i_\mu(p)$ ($\widehat J^{L\,i}_\mu(p)$) the Fourier transform of $W^i_\mu(x)$ ($J^{L\,i}_\mu(x)$). The leftmost diagram of  the top panel of fig.~\ref{fig:FIG33} was already computed (see eq.~(\ref{MW})) and contributes the first term in eq.~(\ref{TPF}). To compute the rightmost diagram we need to evaluate in the critical limit the (amputated) $W$-NG boson two-point function as well as the NG-boson propagator (the latter is depicted in the bottom panel of fig.~\ref{fig:FIG33}).

These are pretty straightforward calculations which do not differ much from the analogous ones we would do in the SM~\footnote{For instance, as in the SM, from eq.~(\ref{TPF}) we can read off the value of the squared $W$ mass (see eq.~(20.11) of~\cite{PESK}).}. The only difference is that here we need to remember that these two-point functions come completely from NP effects, as the perturbative contribution (recall eq.~(\ref{KCR1}))
\beq 
\Gamma^{PT}_{\rm kin}(\zeta)=\frac{v^2}{F^2}\Big{[}k_{b} - (\rho_q^2 N_ck_{bq}^{(1)} + \rho_Q^2 N_c  N_T k_{bQ}^{(1)})+\ldots\Big{]} \,\partial_\mu \vec \zeta \,\partial_\mu \vec \zeta\, \stackrel {k_b\to k_{b\, cr}} \longrightarrow 0
\, ,\label{L1}
\eeq
vanishes in the critical limit. In eq.~(\ref{L1}) we have used the standard polar decomposition 
\beqn
&&\Phi=R\, U\equiv(v+\zeta_0)U\, , 
\qquad U=\exp[i\vec \tau \vec \zeta/F]\, , \label{PHUF}
\eeqn
where for a while $F$ is left as a not yet specified scale. The vanishing of the quantity in the square parenthesis in eq.~(\ref{L1}) is precisely the condition that determine the critical value of $k_b$ in eq.~(\ref{KCR1}). We recall from sect.~\ref{sec:TCT} that the first contribution in the parenthesis comes from the NG-boson kinetic term present in the fundamental Lagrangian~(\ref{LWILWQ}), the second from the 1-loop fermion correction of fig.~\ref{fig:fig5c} (see eq.~(\ref{KCR1})) and the dots from the rest of the loop expansion. 

From the NP expression of the QEL of the critical theory given in eq.~(\ref{GWQCNG}), we get
\beq
\langle \widehat W^i_\mu(p)\widehat \zeta^j(-p)\rangle \Big{|}^{\rm amp}_{p\to 0} \to c^2_w\Lambda_T^2 g_w\frac{p_\mu}{F} \delta_{ij} 
\label{TPF1}
\eeq 
for the $W$-NG boson two-point function and 
\beq
\langle \widehat \zeta^i(p)\widehat \zeta^j(-p)\rangle \Big{|}^{\rm amp}_{p\to 0}=\frac{c^2_w\Lambda_T^2 }{F^2} \frac{1}{p^2} \delta_{ij} \label{L2}\eeq
for the (tree-level) NG-boson propagator.
\begin{figure}
\includegraphics[scale=0.4]{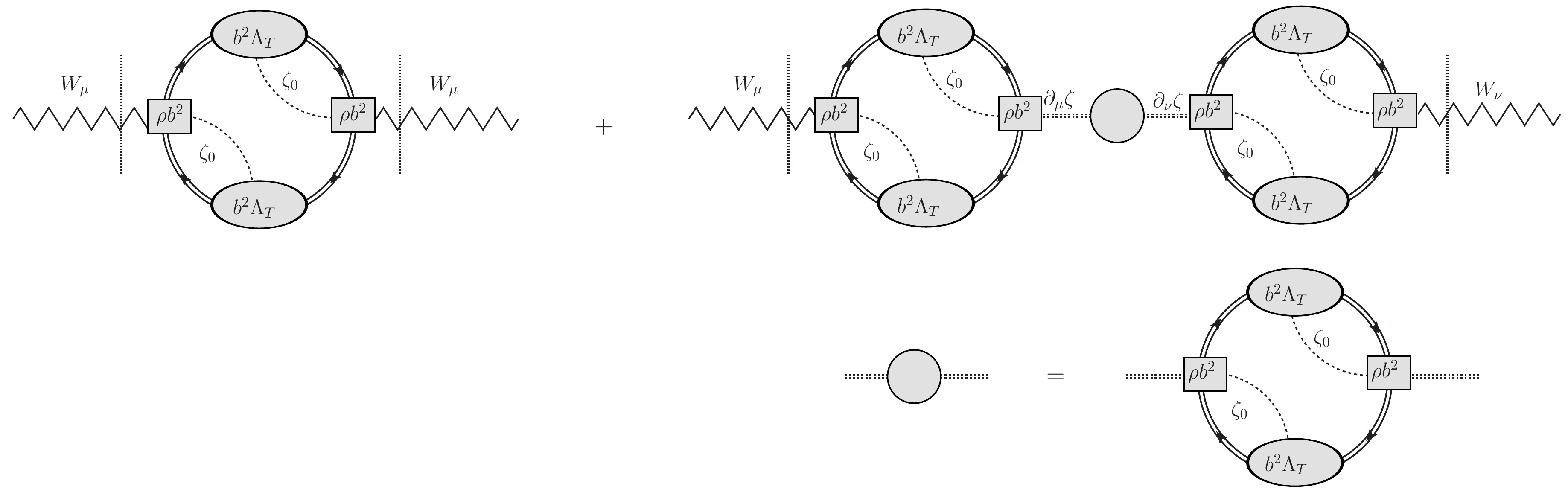} 
\caption{Top panel: the diagrams making transverse the $W$ polarization amplitude. The doubly dotted lines represent the propagation of a NG boson. Bottom panel: the NG-boson propagator. The rest of the notation is as in the figures in sects.~\ref{sec:CCNGP} and~\ref{sec:EEPM}.}  
\label{fig:FIG33}
\end{figure}

Putting everything together, one obtains for the sum of the two diagrams in the top panel of fig.~\ref{fig:FIG33} the desired transverse expression
{\beqn
\hspace{-.8cm}&&g_w^2 \langle \widehat J^{L\,i}_\mu(p) \widehat J^{L\,j}_\nu(-p)\rangle \Big{|}_{p\to 0}\!\!\! \to\! \Big{[} c^2_w\Lambda_T^2 g_w^2 \delta_{\mu\nu}\!-\! \Big{(}c^2_w\Lambda_T^2 g_w\frac{p_\mu}{F} \Big{)}\frac{F^2}{c^2_w\Lambda_T^2} \frac{1}{p^2} \Big{(}\frac{p_\nu}{F} c^2_w\Lambda_T^2g_w\Big{)} \Big{]} \delta_{ij}\!=\nn\\
\hspace{-.8cm}&&\qquad = c^2_w\Lambda_T^2 g_w^2\Big{[} \delta_{\mu\nu} \!-\! \frac{p_\mu p_\nu}{p^2}\Big{]} \delta_{ij} \, .
\label{TWOPW}
\eeqn
It should be stressed that, as expected, the arbitrary scale $F$ introduced in the parametrization~(\ref{PHUF}) completely disappears from the physical formula~(\ref{TWOPW}). Any way, as we have already remarked, in view of the form~(\ref{GWQCNG}) of the QEL it is convenient to set $F=c_w\Lambda_T$ in eq.~(\ref{PHUF}) so as to have canonically normalized Goldstone fields.

{\bf Acknowledgments -} We are indebted to R.\ Frezzotti for his interest in this work and for infinitely many comments and suggestions on the issues presented in this paper. We wish to thank R.\ Barbieri, M.\ Bochicchio, G.\ Martinelli, C.~T.\ Sachrajda, N.\ Tantalo, M.\ Testa and especially G.\ Veneziano for many useful discussions. We thank M.\ Garofalo for a careful reading of the manuscript and for illuminating correspondence.

\end{document}